\documentstyle[psfig]{mn}

\newif\ifAMStwofonts
\AMStwofontstrue

\def\spose#1{\hbox to 0pt{#1\hss}}
\def \da{\Delta\alpha/\alpha}
\def \nalpha{\not {\hspace{0.1mm}\alpha}}
\def \hiz{high-$z$}
\def \loz{low-$z$}

\title[Possible evidence for a variable $\alpha$ from QSO absorption lines:
motivations, analysis and results] {Possible evidence for a variable fine
structure constant from QSO absorption lines: motivations, analysis and
results\thanks{Data presented herein were obtained at the W.M. Keck
Observatory, which is operated as a scientific partnership among the
California Institute of Technology, the University of California and the
National Aeronautics and Space Administration. The Observatory was made
possible by the generous financial support of the W.M. Keck Foundation.}}

\author[M. T. Murphy et al.]  {M. T. Murphy$^1$, J. K. Webb$^1$\thanks{E-mail:
       jkw@bat.phys.unsw.edu.au (JKW)}, V. V. Flambaum$^1$,
       V. A. Dzuba$^1$,\newauthor C. W. Churchill$^2$, J. X. Prochaska$^3$,
       J. D. Barrow$^4$ and A. M. Wolfe$^5$\\
$^1$School of Physics, The
       University of New South Wales, UNSW Sydney NSW 2052, Australia\\
$^2$Department of Astronomy \& Astrophysics, Pennsylvania State
       University, University Park, PA, 16802, USA\\
$^3$The Observatories
       of the Carnegie Institute of Washington, 813 Santa Barbara
       St. Pasadena, CA 91101\\
$^4$Department of Applied Mathematics and Theoretical Physics, Centre for
       Mathematical Sciences, University of Cambridge,\\~Wilberforce Road,
       Cambridge CB3 0WA, England, UK\\
$^5$Department of Physics and Center for
       Astrophysics and Space Sciences, University of California, San
       Diego,\\~C-0424, La Jolla, CA 920923}

\date{Accepted 2001 July 9.
      Received 2001 July 5;
      in original form 2000 December 20}

\pagerange{\pageref{firstpage}--\pageref{lastpage}}
\pubyear{2000}

\begin{document}

\maketitle

\label{firstpage}

\begin{abstract}
An experimental search for variation in the fundamental coupling constants
is strongly motivated by modern high-energy physics theories.  Comparison
of quasar absorption line spectra with laboratory spectra provides a
sensitive probe for variability of the fine structure constant, $\alpha$,
over cosmological time-scales. We have previously developed and applied a
new method providing an order of magnitude gain in precision over previous
optical astrophysical constraints. Here we extend that work by including
new quasar spectra of damped Lyman-$\alpha$ absorption systems.  We also
re-analyse our previous lower redshift data and confirm our initial
results. The constraints on $\alpha$ come from simultaneous fitting of
absorption lines of subsets of the following species: Mg{\sc \,i}, Mg{\sc
\,ii}, Al{\sc \,ii}, Al{\sc \,iii}, Si{\sc \,ii}, Cr{\sc \,ii}, Fe{\sc
\,ii}, Ni{\sc \,ii} and Zn{\sc \,ii}.  We present a detailed description of
our methods and results based on an analysis of 49 quasar absorption
systems (towards 28 QSOs) covering the redshift range $0.5 < z < 3.5$.
There is {\it statistical} evidence for a smaller $\alpha$ at earlier
epochs: $\da = (-0.72 \pm 0.18) \times 10^{-5}$.  The new and original
samples are {\it independent} but separately yield consistent and
significant non-zero values of $\da$.  We summarise the results of a
thorough investigation of systematic effects published in a companion
paper. The value we quote above is the raw value, not corrected for any of
these systematic effects. The only significant systematic effects so far
identified, if removed from our data, would lead to a {\it more
significant} deviation of $\da$ from zero.
\end{abstract}

\section{Introduction}\label{sec:intro}

High resolution spectroscopy of QSO absorption systems provides a precise
probe of fundamental physics.  In previous papers (Dzuba, Flambaum \& Webb
1999a,b; Webb et al. 1999, hereafter W99) we introduced and applied a new
and highly sensitive method for constraining space-time variations of the
fine structure constant, $\alpha \equiv \frac{e^2}{\hbar c}$.  Using 30
Keck/HIRES absorption spectra containing Mg{\sc \,ii} and Fe{\sc \,ii}, we
reported a tentative detection of a variation in $\alpha$: $\da = (-1.09
\pm 0.36) \times 10^{-5}$, where $\da$ is defined as
\begin{equation}
\da = (\alpha_z - \alpha_0)/\alpha_0,
\end{equation}
$\alpha_0$ is the present day value of $\alpha$ and $\alpha_z$ is the value
at the absorption redshift, $z$. The absorption redshifts ranged from
$\sim$0.6--1.6.

Here we continue this work, presenting new measurements of $\da$ in high
redshift damped Lyman-$\alpha$ systems (DLAs) over the range $0.9 < z <
3.5$, and a re-analysis of the W99 data set. Four papers present our new
work on this subject.  A brief summary of all results is given in Webb et
al. (2001).  The present paper provides a detailed description of our
methods and results. A companion paper (Murphy et al. 2001b, hereafter
M01b) presents a thorough examination of systematic effects.  An additional
paper (Murphy et al. 2001c) presents an analysis of a sample of Si{\sc
\,iv} QSO absorbers.

The present paper is organized as follows. Section \ref{sec:previousth}
describes theoretical motivations for a search for varying $\alpha$. In
Section \ref{sec:previousex} we summarise previous experimental constraints
on varying $\alpha$.  Section \ref{sec:theory} presents the theoretical
basis of the new method for determining $\da$ using QSO absorbers.  Section
\ref{sec:req} describes the experimental components of our study:
high-resolution QSO observations and precise laboratory rest wavelength
measurements.  In Section \ref{sec:anal} we discuss our analysis technique
in detail and present our results. For completeness we also summarise the
systematic effects discussed in detail in M01b.  Section
\ref{sec:discussion} summarises our new results and outlines future work.

\section{Theoretical background}\label{sec:previousth}

\subsection{Early historical interest}\label{sec:history}

The constancy of the fundamental constants began to be questioned when
Milne (1935, 1937) and Dirac (1937) suggested that the Newtonian
gravitational constant, $G$, may vary in cosmological time. Milne proposed
a gravitational theory in which different physical `clocks' -- `ticking'
according to different physical processes -- `tick' at different rates as
time progresses. The possible consequences of this hypothesis for
biological and geological history were seized upon by Haldane (1937a,b) to
explain aspects of their discontinuous development, whereas Kothari (1938)
and Chandrasekhar (1939) pointed out some astronomical consequences of
Dirac's 'large numbers hypothesis' which might permit it to be tested by
observation. Dirac's Large Numbers Hypothesis attracted considerable
interest (e.g. Teller 1948; Dicke 1957, 1961; Barrow 1990; Barrow \& Tipler
1996) and the desire to place his idea of varying $G$ on a firmer
theoretical footing led to the development of scalar-tensor generalisations
of Einstein's theory of general relativity in which the variation of $G$
was described self-consistently by the propagation of a scalar field which
also acted as a source of space-time curvature (Brans \& Dicke 1961).

Jordan (1937, 1939) was the first to consider how Dirac's hypothesis might
be applied to forces other than gravity, but rejected the possibility of
time variations in the weak interaction or the electron-proton mass ratio
(Jordan 1937, 1939). Later, Gamow (1967) showed that geological problems
caused by Dirac's proposed variation in $G$ that had been pointed out by
Teller (1948) could be avoided by interpreting the large numbers hypothesis
as requiring the time-variation of the charge on the electron, $e$, rather
than $G$. Observational limits on possible variations of $\alpha$ or the
weak and strong interactions were more restrictive than those on $G$
although no self-consistent theory permitting the variation of
non-gravitational force constants was developed to make the observational
limits rigorous. Landau (1955) proposed that variation in $\alpha$ could be
connected to renormalization rules in quantum electrodynamics.

Current motivation for a search for a varying $\alpha$ comes from modern
theories which attempt to unify gravity with other interactions.

\subsection{Modern theoretical motivations for varying $\alpha$}\label{sec:motiv}

At present there are many theoretical motivations for a search for varying
$\alpha$. However, these theories, being relatively recent, may not be
widely appreciated within the context of astronomical tests of their
viability. We therefore summarise recent ideas below.

\subsubsection{Multi-dimensional unification theories}\label{sec:unified}
One of the greatest challenges facing modern theoretical physics is the
quantization of the gravitational interaction. Early attempts at creating
such a unification were made using a geometrization in ($4+D$)--dimensional
curved space-times in the spirit of the old Kaluza-Klein scenario to unite
gravity and electromagnetism (Kaluza 1921; Klein 1926).  Three-dimensional
gauge couplings like $\alpha$, $\alpha_{\rm weak}$, and $\alpha_{\rm
strong}$ then must vary as the inverse square of the mean scale of the
extra dimensions. Thus, the evolution of the scale size of the extra
dimensions is related to variability of the low-energy coupling constants
in the 4--dimensional subspace in simple Kaluza-Klein and superstring
theories.

Damour \& Polyakov (1994) showed that cosmological variation in $\alpha$
may proceed at different rates at different points in space-time (see also
Forg\'{a}cs \& Horv\'{a}th 1979; Barrow 1987; Li \& Gott 1998). Various
functional forms for time variation of $\alpha/G$ have been derived using
Kaluza-Klein theory and the assumption of constant masses. Chodos \&
Detweiler (1980) find $\alpha/G\propto t$ and Freund (1982) finds
$\alpha/G\propto t^{1/4}$. Wu \& Wang (1986) have found that $\frac{\dot
\alpha }\alpha \propto \frac{\dot G}G\sim -10^{-11\pm 1}{\rm yr}^{-1}$
where the proportionality constant is unknown.  It is interesting to note
that using a similar analysis to Wu \& Wang, Maeda (1988) derived $\alpha
\propto t^{-4/3}$.

Marciano (1984) discussed the self-consistency relations required if there
are simultaneous variations of different constants in unified gauge
theories.  He also examined any possible non-monotonic variation in
$\alpha$ with $t$, using the running coupling dependence of strong, weak,
and electromagnetic interactions to produce self-consistent predictions for
the simultaneous variation of more than one coupling or mass ratio.  These
were discussed in Drinkwater et al. (1998). Typically, variation in $G$ and
$\alpha$ could be linked by relations of the form
$\frac{\Delta\alpha}{\alpha^2}\sim \frac{\Delta G}{G}$.

These features of extra dimensions mean that in all string theories (and
M-theory, of which they are presumed to be limiting cases, Ho\v{r}ava \&
Witten 1996a,b), any extra dimensions of space need to be held static with
high accuracy in order to avoid gross conflict with observation. In the
currently popular scenarios for M-theory (e.g. Antoniadis et al. 1998;
Arkani-Hamed, Dimopoulos \& Dvali 1998; Randall \& Sundrum 1999a,b), only
the gravitational force is assumed to act in all ($>3$) spatial dimensions
(the `bulk') whilst all other interactions act only in 3-dimensional space
(the `brane'). Observations of the constancy of 3-dimensional
non-gravitational constants governing the strengths of those interactions
in three dimensions (like $\alpha$) could therefore be of crucial
importance in testing this theoretical scenario. The scale of the extra
dimensions was until recently believed to be necessarily of order the
Planck scale $\sim 10^{-33}{\rm \,cm}$ but it is possible they could be as
large as $\sim 0.01{\rm \,mm}$ leading to large changes in the form of the
Newtonian law of gravity on length scales below this distance. At present,
this is consistent with all known experiments. The challenge of testing the
law of gravity on sub-millimetre scales is considerable because other
forces strongly dominate the effects of gravity on these scales.

\subsubsection{Scalar Theories}
The first self consistent theory of electromagnetism which incorporates
varying $\alpha$, and which reduces to Maxwell's theory in the limit of
constant $\alpha$, was that developed by Bekenstein (1982). It introduces a
scalar field whose variation in space and time produces effective variation
in the electron charge (or, equivalently, the permittivity of free space).
The form of the propagation equation for the evolution of the scalar field
is strongly constrained by the natural requirements that it be second
order, causal and linearly coupled to the matter density. This particular
theory has been constrained by several astronomical observations (Livio \&
Stiavelli 1998) and used to illustrate consequences of space and time
variations of constants (Barrow \& O'Toole 2000).

The general structure of Bekenstein's theory can also be applied to other
couplings, and it has recently been used to investigate the problem of
varying the strong coupling constant of QCD by Chamoun, Landau \& Vucetich
(2000).  Bekenstein makes the simplifying assumption that the gravitational
sector of the theory be unchanged from that of general relativity. This
amounts to neglecting the contribution of the kinetic energy of the scalar
field associated with $\alpha$ variation to the expansion dynamics of the
universe. This defect can be avoided in other versions of this theory
(Barrow \& Magueijo 2000; Magueijo 2000). It should be emphasised that
there is no single self-consistent theory incorporating varying $\alpha$
and theoretical limits need to be quoted in conjunction with the
theoretical framework used to confront observational data. If different
interactions are unified by electro-weak or grand unified theories then the
ramifications of introducing varying $\alpha$ will be deeper and wider and
a scalar theory would need to accommodate the effects of varying $\alpha$
on the other gauge couplings.

\subsubsection{Varying speed of light theories}\label{sec:vsl}
One appeal of varying speed of light theories is that they provide
alternative potential solutions to a range of cosmological problems such as
the flatness, horizon and monopole problems (Moffat 1993; Barrow \&
Magueijo 1999a,b; Albrecht \& Magueijo 1999; Barrow 1999).  Barrow \&
Magueijo (1998) have also expressed Bekenstein's (1982) varying-$e$ theory
in terms of varying $c$.

Barrow \& Magueijo (2000) developed a particular theory for varying $c$ (or
$\alpha$) in which the stress contributed by the cosmological constant
varies through the combination $\Lambda c^2$.  They also showed how the
observed non-zero cosmological acceleration (Schmidt et al. 1998;
Perlmutter et al. 1999) might be linked to a varying $\alpha$. They
examined a class of varying-$c$ theories in which changes in $c$ are driven
by a scalar field which is coupled to the gravitational effect of
pressure. Changes in $c$ convert the $\Lambda$ energy density into
radiation, preventing $\Lambda$ from dominating the universe in the
radiation epoch. As pressureless matter begins to dominate, variations in
$c$ slow, and $\Lambda$ begins to dominate. The very slow variation of the
scalar field driving changes in $c$ is like that of scalar fields in
quintessence theories in the presence of perfect fluids. This type of
theory allows the time variations in $c$ or $\alpha$ to be $\sim
10^{-5}\,H_0$ at $z\sim 1$ and yet the associated $\Lambda $ term can be
dominant today and produce acceleration (the $10^{-5}$ factor is created
naturally in this theory by the ratio of the radiation to matter density in
the universe today). We therefore expect only very small residual variation
in $\alpha$ in a $\Lambda$ dominated universe. The existence of simple
theories of this sort provides a strong motivation for studies like the
present one.

\subsubsection{Other theories}\label{sec:others} 
Several authors have suggested other motivations for varying $\alpha$. For
example, Hill \& Ross (1988a,b) (see also Hill, Steinhardt \& Turner 1990)
have suggested that oscillatory variations in $\alpha$ might be expected if
there exists a soft boson field that has electromagnetic couplings to
either the Maxwell scalar, $F_{\mu\nu}F^{\mu \nu}$, or to the electron
state, $\bar{\psi_e}\psi_e$. If the boson field has a mass $m$ then the
frequency of emission and absorption of radiation will vary with a
frequency $\sim\,m^{-1}$. An oscillatory $\alpha$ might also be
attributable to the decaying oscillations of a scalar field that couples to
$F_{\mu\nu}F^{\mu\nu}$. Such a theory is allowed by enforcing an
approximate global symmetry (Carroll 1998).

\bigskip

In summary, there is much current theoretical motivation for experimental
searches for variability in $\alpha$, particularly at medium to high
redshifts. A detection of variation may allow us to probe the extra
dimensions expected from many modern unified theories. A search for varying
$\alpha$ provides an important tool for seeking out and constraining the
form of new physics.

\section{Previous experimental work}\label{sec:previousex}

\subsection{Terrestrial and laboratory constraints}\label{sec:terre}
Direct laboratory measurements provide interesting constraints on a
time-varying $\alpha$.  Prestage, Robert \& Maleki (1995) introduced a new
technique comparing the rates of clocks based on ultra-stable oscillators.
Relativistic corrections are order $(Z_n\alpha)^2$ where $Z_n$ is the
atomic number or nuclear charge.  By comparing the rates of two clocks
based on different atoms (H-maser and Hg$^+$) over a 140 day period, they
were able to constrain $\left|\dot\alpha/\alpha\right| \le 3.7 \times
10^{-14}{\rm \,yr}^{-1}$ (i.e. $\left|\da\right| \le 1.4 \times 10^{-14}$).

Model dependent upper limits on any variation have been claimed from an
analysis of the Oklo phenomenon -- a natural nuclear fission reactor that
operated in Gabon, West Africa, $\sim$1.8 billion years ago.  Shlyakhter
(1976) originally constrained the energy of the nuclear resonance level in
the $^{150}$Sm isotope and related this to upper bounds on
$\left|\da\right|$. Damour \& Dyson (1996) have also analysed the problem
with different model assumptions and claim $-0.9 \times 10^{-7} < \da < 1.2
\times 10^{-7}$. Recently, Fujii et al. (2000) obtained somewhat tighter
constraints using new samples from the Oklo reactor: $\da = (-0.04\pm
0.15)\times 10^{-7}$.

However, the Oklo limit corresponds to variations at very low `redshift',
$z \sim 0.1$, in a non-cosmological environment.  In the absence of a
detailed theory giving both time {\it and} space variations of $\alpha$, it
is dangerous to compare time variations in areas of different gravitational
potential (Barrow \& O'Toole 2000).  Also, there are considerable
uncertainties regarding the complicated Oklo environment and any limits on
$\alpha$ derived from it can be weakened by allowing other interaction
strengths and mass ratios to vary in time as well.

\subsection{Constraints from QSO spectra}\label{sec:qso}
Savedoff (1956) first analysed doublet separations seen in galaxy emission
spectra to constrain possible time variation of $\alpha$.  Absorption lines
in intervening clouds along the line of sight to QSOs are substantially
narrower than intrinsic emission lines and therefore provide a more precise
probe of $\alpha$. This prompted Bahcall, Sargent \& Schmidt (1967) to use
alkali-doublet spacings seen in the absorption spectra of QSOs to arrive at
firm upper limits on any variation.  The appeal of this method is the huge
time span probed by high-redshift observations.  Wolfe, Brown \& Roberts
(1976) first applied the alkali-doublet (AD) method to intervening Mg{\sc
\,ii} absorption.  Varshalovich, Panchuk \& Ivanchik (1996) have used the
AD method for many absorbers to obtain upper limits on any variation at
redshifts $z \sim 2.8$--$3.1$: $\da = (2 \pm 7) \times 10^{-5}$.

Varshalovich, Potekhin \& Ivanchik (2000) have recently analysed 16 Si{\sc
\,iv} absorption systems (towards 6 QSOs) using the AD method. They find a
mean $\da = (-4.6 \pm 4.3 \pm 1.4) \times 10^{-5}$ using the Si{\sc \,iv}
laboratory wavelengths of Morton (1991).  The systematic component to the
error above ($1.4 \times 10^{-5}$) allows for the relative uncertainties of
the Si{\sc \,iv} laboratory wavelengths.  However, Griesmann \& Kling
(2000) have increased the absolute precision of these wavelengths and their
new values indicate that the Varshalovich et al. (2000) result should be
corrected by $\approx +11.3 \times 10^{-5}$ to $\da \approx (+7 \pm 4)
\times 10^{-5}$ suggesting that the systematic error component added by
Varshalovich et al. was significantly underestimated.

In Murphy et al. (2001c) we have used the new Griesmann \& Kling laboratory
wavelengths to constrain $\da$ using 21 Si{\sc \,iv} doublets towards 8
QSOs.  The spectral resolution and SNR are higher than for the Varshalovich
et al. sample.  This, and the higher precision laboratory wavelengths,
provide a more stringent constraint of $\da = (-0.5 \pm 1.3) \times
10^{-5}$.  This result is nevertheless a factor of 7 less precise than the
results we present in this paper, highlighting the strength of the methods
we use.

The optical techniques above were based on the AD method: constraining
$\alpha$ from the splitting of single ADs such as Si{\sc \,iv}, C{\sc \,iv}
and Mg{\sc \,ii}. Very recently, however, Dzuba et al. (1999a) and W99
proposed a new method that allows {\it an order of magnitude increase in
precision}. The method is based on measuring the wavelength separation
between the resonance transitions of {\it different ionic species} with no
restriction on the multiplet to which the transitions belong. We outline
this many-multiplet (MM) method and discuss its advantages in Section
\ref{sec:theory} of this paper.

Radio spectra of QSO absorption clouds also provide the opportunity for
tightly constraining possible variations in the fundamental
constants. Wolfe et al. (1976) demonstrated that combining optical
resonance and H{\sc \,i}\,$21{\rm \,cm}$ hyperfine absorption spectra could
yield constraints on $\alpha^2g_p\frac{m_e}{m_p}$. Here, $g_p$ is the
proton g-factor and $m_e$ and $m_p$ are the mass of the electron and proton
respectively; see also the discussion by Pagel (1983).  Varshalovich \&
Potekhin (1996) used a comparison of H{\sc \,i}\,$21{\rm \,cm}$ hyperfine
and molecular rotational absorption to constrain variations in $m_p$ but
Drinkwater et al. (1998) showed that such a combination actually constrains
$y\equiv\alpha^2g_p$. In Murphy et al. (2001d) we have also made a more
complete analysis of the two Drinkwater et al. radio spectra to obtain
constraints on $y$. Assuming $g_p$ to be constant, we find $\da = (-0.10\pm
0.22) \times 10^{-5}$ and $\da = (-0.08 \pm 0.27) \times 10^{-5}$ at
$z=0.25$ and $0.68$ respectively. If we note the \loz~points in the lower
panel (binned data) of Fig. 3 then we see that our results are also
consistent with the two radio points. Carilli et al. (2000) have also
recently analysed independent spectra of these two QSOs and find an upper
limit on any variation consistent with our measurements: $\left|\da\right|
< 0.85 \times 10^{-5}$ (again assuming $g_p$ to be constant).

Cowie \& Songaila (1995) have also obtained constraints at $z \approx
1.78$. They compare the observed frequencies of the H{\sc \,i}\,$21{\rm
\,cm}$ and UV C{\sc \,i} absorption lines which are sensitive to variations
in $g_p\alpha^2\frac{m_e}{m_p}$. If one assumes that $g_p\frac{m_e}{m_p}$
is constant then their constraint corresponds to $\da = (0.35 \pm 0.55)
\times 10^{-5}$.

\subsection{Early universe (CMB and BBN)}\label{sec:early}
A smaller value of $\alpha$ in the past would change the ionization history
of the universe, postponing the recombination of electrons and protons,
i.e. last-scattering would occur at lower redshift.  It would also alter
the ratio of baryons to photons at last-scattering, leading to changes in
both the amplitudes and positions of features in the cosmic microwave
background (CMB) power spectrum, primarily at angular scales $\la
1^{\circ}$ (Hannestad 1999; Kaplinghat, Scherrer \& Turner 1999; Avelino,
Martins \& Rocha 2000; Battye, Crittenden \& Weller 2000; Kujat \& Scherrer
2000).

Kolb, Perry \& Walker (1986) have discussed how limits can be placed on
$\da$ at the time of big bang nucleosynthesis (BBN) (corresponding to a
redshift $z \sim 10^8 - 10^9$) by assuming a simple scaling between the
value of $\alpha$ and the neutron-proton mass difference.  Using
observations of the $^4$He abundance they derive limits of
$\left|\dot\alpha/\alpha\right| < 15\times 10^{-15}h_{100}{\rm \,yr}^{-1}$,
corresponding to $\left|\da\right| < 9.9 \times 10^{-5}$.

Barrow (1987) and Campbell \& Olive (1995) considered the simultaneous
variation of weak, strong and electromagnetic couplings on primordial
nucleosynthesis. However, these limits are beset by a crucial uncertainty
as to the electromagnetic contributions to the neutron-proton mass
difference. Much weaker limits are possible if attention is restricted to
the nuclear interaction effects on the nucleosynthesis of deuterium,
helium-3, and lithium-7.

Bergstr\"{o}m, Iguri \& Rubinstein (1999) have re-examined all light elements
up to $^7$Li.  By including elements heavier than $^4$He they avoid the
problem of the unknown electromagnetic contributions to the neutron-proton
mass difference, and argue that a more realistic limit, given the present
observational uncertainties in the light element abundances is more like
$2\times 10^{-2}$.

\section{Theoretical basis of the many-multiplet method}\label{sec:theory}
The MM method allows an order of magnitude greater precision compared to
the previous AD method.  The full details behind this technique are
presented in detail in Dzuba et al. (1999a,b, 2001) and W99 and so we
present only the salient features here.

To explain the MM method, we begin with a simple analytic approach
to estimate the relativistic effects in transition frequencies. If we
consider a many-electron atom/ion then the relativistic correction,
$\Delta$, to the energy of the external electron can be written as
\begin{equation}\label{eq:relcor}
\Delta \propto
(Z_n\alpha)^2\left|E\right|^{3/2}\left[\frac{1}{j+1/2}-C(j,l)\right]\,,
\end{equation}
where $Z_n$ is the nuclear charge, $E$ is the electron energy ($E<0$,
$\left|E\right|$ is the ionization potential) and $j$ and $l$ are the total
and orbital electron angular momenta. The contribution to the relativistic
correction from many-body effects is described by $C(j,l)$. For $s$ and $p$
orbitals, $C(j,l) \approx 0.6$ and is of similar magnitude for $d$
orbitals.  Equation \ref{eq:relcor} therefore provides a general strategy
for probing the relativistic corrections in resonance transitions.

For example, consider comparison of the transition energies of two $s$--$p$
transitions, one in a light ion, the other in a heavy ion (i.e. low and
high $Z_n$ respectively). The $Z_n^2$ term dominates so the relativistic
corrections to the transition energies will differ greatly. Thus,
comparison of the spectra of light and heavy species is a sensitive probe
of $\alpha$.

As a further example, consider an $s$--$p$ and a $d$--$p$ transition in a
heavy species. The corrections will be large in each case but will be of
opposite sign since the many-body corrections, $C(j,l)$, begin to dominate
with increasing $j$. This situation also allows tight constraints to be
placed on $\alpha$.

Thus, comparing spectra of transitions from different multiplets and
different atoms or ions, provides a sensitive method for probing variations
in $\alpha$. In comparison, the fine splitting of an $s$--$p$ doublet will
be substantially smaller than the absolute shift in the $s$--$p$ transition
energy since the excited $p$ electron, with relatively small
$\left|E\right|$, will have much smaller relativistic corrections than the
$s$ electron. Therefore, the AD method is relatively insensitive to
variations in $\alpha$.

\begin{table*}
\caption{Atomic data for those transitions used in our analysis. The
origins of the composite wavenumbers ($\omega_0$) and wavelengths
($\lambda_0$) are detailed in Section \ref{sec:lab}. We give the ground and
excited state configurations in columns 4 and 5. The ID letters in column 6
are used in Table \ref{tab:results} to indicate the transitions used in our
Voigt profile fits to each absorption system. The seventh column shows the
ionization potential (IP)$^a$ and the eighth shows the literature
oscillator strengths, $f$. Values of $q_1$ and $q_2$ are also shown. The
Si{\sc \,ii}, Al{\sc \,ii} and Al{\sc \,iii} wavenumbers have been scaled
from their literature values due to the Norl\'{e}n/Whaling et
al. calibration difference described in Section \ref{sec:lab}. The isotopic
structures for the Mg and Si transitions are given in Table \ref{tab:iso}.}
\begin{minipage}{177mm}
\label{tab:atomdata}
\begin{center}
\begin{tabular}{lllllcrlrr}\hline
\multicolumn{1}{c}{Ion}&\multicolumn{1}{c}{$\lambda_0$}&\multicolumn{1}{c}{$\omega_0$}&\multicolumn{1}{c}{Ground}&\multicolumn{1}{c}{Upper}&\multicolumn{1}{c}{ID}&\multicolumn{1}{c}{IP}&\multicolumn{1}{c}{$f$}&\multicolumn{1}{c}{$q_1$}&\multicolumn{1}{c}{$q_2$}\\
&\multicolumn{1}{c}{$\overline{{\rm
\AA}}$}&\multicolumn{1}{c}{$\overline{{\rm
cm}^{-1}}$}&\multicolumn{1}{c}{state}&\multicolumn{1}{c}{state}&&\multicolumn{1}{c}{$\overline{{\rm
eV}}$}&&\multicolumn{1}{c}{$\overline{{\rm
cm}^{-1}}$}&\multicolumn{1}{c}{$\overline{{\rm cm}^{-1}}$}\\\hline
Mg{\sc \,i}  &2852.96310(8)&35051.277(1)$^b$ &$3s^2~^1{\rm S}_0         $&$3s3p~^1{\rm P}_1         $&a &  -- &1.810$^g$  &  106&-10\\ 
Mg{\sc \,ii} &2796.3543(2) &35760.848(2)$^b$ &$3s~^2{\rm S}_{1/2}       $&$3p~^2{\rm P}_{3/2}       $&b & 7.7 &0.6295$^h$ &  211&  0\\
             &2803.5315(2) &35669.298(2)$^b$ &$                         $&$3p~^2{\rm P}_{1/2}       $&c &     &0.3083$^h$ &  120&  0\\
Al{\sc \,ii} &1670.7887(1) &59851.972(4)$^c$ &$3s^2~^1{\rm S}_0         $&$3s3p~^1{\rm P}_1         $&d & 6.0 &1.88$^g$   &  270&  0\\
Al{\sc \,iii}&1854.71841(3)&53916.540(1)$^c$ &$3s~^2{\rm S}_{1/2}       $&$3p~^2{\rm P}_{3/2}       $&e &18.9 &0.268$^g$  &  464&  0\\
             &1862.79126(7)&53682.880(2)$^c$ &$                         $&$3p~^2{\rm P}_{1/2}       $&f &     &0.539$^g$  &  216&  0\\
Si{\sc \,ii} &1526.70709(2)&65500.4492(7)$^c$&$3s^23p~^2{\rm P}_{1/2}^o $&$3s^24s~^2{\rm S}_{1/2}   $&g & 8.2 &0.116$^i$  &   24& 22\\
             &1808.01301(1)&55309.3365(4)$^c$&$                         $&$3s3p^2~^2{\rm D}_{3/2}   $&h &     &0.00218$^g$&  525&  3\\
Cr{\sc \,ii} &2056.25693(8)&48632.055(2)$^d$ &$3d^5~^6{\rm S}_{5/2}     $&$3d^44p~^6{\rm P}_{7/2}   $&i & 6.8 &0.105$^j$  &-1030&-13\\
             &2062.23610(8)&48491.053(2)$^d$ &$                         $&$3d^44p~^6{\rm P}_{5/2}   $&j &     &0.078$^j$  &-1168&-16\\
             &2066.16403(8)&48398.868(2)$^d$ &$                         $&$3d^44p~^6{\rm P}_{3/2}   $&k &     &0.0515$^j$ &-1267& -9\\
Fe{\sc \,ii} &1608.45085(8)&62171.625(3)$^e$ &$3d^64s~a^6{\rm D}_{9/2}  $&$3d^64p~y^6{\rm P}_{7/2}  $&l & 7.9 &0.0619$^k$ & 1002&141\\
             &1611.20034(8)&62065.528(3)$^e$ &$                         $&$3d^64p~y^4{\rm F}_{7/2}  $&m &     &0.00102$^k$& 1110& 48\\
             &2344.2130(1) &42658.2404(2)$^f$&$                         $&$3d^64p~z^6{\rm P}_{7/2}  $&n &     &0.110$^k$  & 1325& 47\\
             &2374.4603(1) &42114.8329(2)$^f$&$                         $&$3d^64p~z^6{\rm F}_{9/2}  $&o &     &0.0326$^k$ & 1730& 26\\
             &2382.7642(1) &41968.0642(2)$^f$&$                         $&$3d^64p~z^6{\rm F}_{11/2} $&p &     &0.300$^k$  & 1580& 29\\
             &2586.6496(1) &38660.0494(2)$^f$&$                         $&$3d^64p~z^6{\rm D}_{7/2}^o$&q &     &0.0684$^k$ & 1687&-36\\
             &2600.1725(1) &38458.9871(2)$^f$&$                         $&$3d^64p~z^6{\rm D}_{9/2}^o $&r &     &0.213$^k$  & 1449&  2\\
Ni{\sc \,ii} &1709.6042(1) &58493.071(4)$^d$ &$3d^9~^2{\rm D}_{5/2}     $&$3d^84p~z^2{\rm F}_{5/2}  $&s & 7.6 &0.0348$^l$ &  800&  0\\
             &1741.5531(1) &57420.013(4)$^d$ &$                         $&$3d^84p~z^2{\rm D}_{5/2}  $&t &     &0.0419$^l$ & -700&  0\\
             &1751.9157(1) &57080.373(4)$^d$ &$                         $&$3d^84p~z^2{\rm F}_{7/2}  $&u &     &0.0264$^l$ & -300&  0\\
Zn{\sc \,ii} &2026.13709(8)&49355.002(2)$^d$ &$3d^{10}4s~^2{\rm S}_{1/2}$&$3d^{10}4p~^2{\rm P}_{3/2}$&v & 9.4 &0.489$^j$  & 2291& 94\\
             &2062.66045(9)&48481.077(2)$^d$ &$                         $&$3d^{10}4p~^2{\rm P}_{3/2}$&w &     &0.256$^j$  & 1445& 66\\\hline
\end{tabular}
\end{center}
{\footnotesize $^a$IP is defined here to be the energy required to
form the ion in question from the ion with a unit lower charge;
$^b$Pickering, Thorne \& Webb (1998); $^c$Griesmann \& Kling (2000);
$^d$Pickering et al. (2000); $^e$S. Johansson (private communication);
$^f$Nave et al. (1991); $^g$Morton (1991); $^h$Verner, Verner \&
Ferland (1996); $^i$Dufton et al. (1983); $^j$Bergeson \& Lawler
(1993); $^k$Cardelli \& Savage (1995); $^l$Fedchak \& Lawler (1999)}
\end{minipage}
\end{table*}

More formally, the energy equation for a transition from the ground state,
within a particular multiplet, at a redshift $z$ can be written as
\begin{equation}\label{eq:energy}
\begin{array}{ll}
E_z=&{\displaystyle \hspace{-2mm}E_{\rm c} + Q_1Z_n^2\left [
\left(\frac{\alpha_z}{\alpha_0}\right)^2-1\right]+}\\
&{\displaystyle K_1({\bf LS})Z_n^2\left(\frac{\alpha_z}{\alpha_0}\right)^2 +
K_2({\bf LS})^2Z_n^4\left(\frac{\alpha_z}{\alpha_0}\right)^4}\,,
\end{array}
\end{equation}
where $\alpha_z$ may or may not be equal to the laboratory value,
$\alpha_0$.  Here, {\bf L} and {\bf S} are the electron total orbital
angular momentum and total spin respectively and $E_{\rm c}$ is the energy
of the configuration centre. $Q_1$, $K_1$ and $K_2$ are relativistic
coefficients which have been accurately computed in Dzuba et al. (1999a,b,
2001) using {\it ab initio} many-body calculations. These calculations were
tested by comparison with experimental data for energy levels and fine
structure intervals. Equation \ref{eq:energy} forms the basis of the MM
method since it can be applied to any resonance transition observed in the
QSO spectra -- another advantage it has over the AD method.

For our purposes, the most convenient form of equation \ref{eq:energy} is
written as
\begin{equation}\label{eq:omega}
\omega_z = \omega_0 + q_1x + q_2y\,,
\end{equation}
where $\omega_z$ is the wavenumber in the rest-frame of the cloud, at
redshift $z$, $\omega_0$ is the wavenumber as measured on Earth and $x$ and
$y$ contain the information about a possible non-zero $\da$:
\begin{equation}
x\equiv\left(\frac{\alpha_z}{\alpha_0}\right)^2-1
~~~{\rm and}~~~
y\equiv\left(\frac{\alpha_z}{\alpha_0}\right)^4-1. 
\end{equation}
The values of the relativistic corrections, $q_1$ and $q_2$, are presented
in Table \ref{tab:atomdata} for all transitions commonly seen in optical
QSO spectra. The values of the $q_1$ coefficients are typically an order of
magnitude larger than the $q_2$ coefficients. Therefore, it is the relative
magnitudes of $q_1$ for different transitions that characterize our ability
to constrain $\da$. The uncertainty in the $q_1$ co-efficients varies from
ion-to-ion and is $\sim 5{\rm \,cm}^{-1}$ for alkali-like ions such as
Mg{\sc \,ii}. However, the co-efficients for the compilcated Ni{\sc \,ii}
transitions carry a larger uncertainty $\sim 200{\rm \,cm}^{-1}$ (Dzuba et
al. 2001). The form of equation \ref{eq:omega} is very convenient since the
second and third terms only contribute if $\da \neq 0$: errors in the $q_1$
and $q_2$ coefficients cannot lead to an artificial non-zero value of
$\da$.

From equation \ref{eq:energy} we can see that $q_1$ will be much larger in
heavier ions due to the $Z_n^2$ dependence. Indeed, one can see in Table
\ref{tab:atomdata} several {\it combinations of transitions from different
ions}, of very different mass, showing large differences between their
respective $q_1$ coefficients. The combination of the Mg{\sc \,i}
$\lambda2853$ line, the Mg{\sc \,ii} doublet and the five Fe{\sc \,ii}
lines between $\lambda2344$ and $\lambda2600$ was used in W99 since the
Fe{\sc \,ii} $q_1$ coefficients typically differ from those of the Mg lines
by almost an order of magnitude. It is this comparison of spectra from
different ions that allows the order of magnitude increase in precision
over previous methods. For comparison, we present the line shifts in
wavenumber, wavelength and velocity space arising from a shift of $\da =
+10^{-5}$ in Table \ref{tab:shifts}.

\begin{table}
\begin{center}
\caption{Shift in the rest frame wavenumber, wavelength and velocity space
for $\da=+10^{-5}$}
\label{tab:shifts}
\begin{tabular}{lrrr}\hline
\multicolumn{1}{c}{Transition}&\multicolumn{1}{c}{$\Delta\omega$}&\multicolumn{1}{c}{$\Delta\lambda$}&\multicolumn{1}{c}{$\Delta v$}\\
&\multicolumn{1}{c}{$\overline{10^{-2}{\rm
cm}^{-1}}$}&\multicolumn{1}{c}{$\overline{10^{-3}{\rm
\AA}}$}&\multicolumn{1}{c}{$\overline{{\rm kms}^{-1}}$}\\\hline
Mg{\sc \,i}\,$\lambda$2853  & 0.17&-0.14 &-0.015 \\      
Mg{\sc \,ii}\,$\lambda$2796 & 0.42&-0.33 &-0.035 \\      
Mg{\sc \,ii}\,$\lambda$2803 & 0.24&-0.19 &-0.020 \\      
Al{\sc \,ii}\,$\lambda$1670 & 0.54&-0.15 &-0.027 \\      
Al{\sc \,iii}\,$\lambda$1854& 0.93&-0.32 &-0.052 \\      
Al{\sc \,iii}\,$\lambda$1862& 0.43&-0.15 &-0.024 \\      
Si{\sc \,ii}\,$\lambda$1526 & 0.14&-0.032&-0.0062\\      
Si{\sc \,ii}\,$\lambda$1808 & 1.06&-0.35 &-0.058 \\      
Cr{\sc \,ii}\,$\lambda$2056 &-2.11& 0.89 & 0.130 \\      
Cr{\sc \,ii}\,$\lambda$2062 &-2.40& 1.02 & 0.148 \\      
Cr{\sc \,ii}\,$\lambda$2066 &-2.57& 1.10 & 0.159 \\      
Fe{\sc \,ii}\,$\lambda$1608 & 2.57&-0.66 &-0.124 \\      
Fe{\sc \,ii}\,$\lambda$1611 & 2.41&-0.63 &-0.117 \\      
Fe{\sc \,ii}\,$\lambda$2344 & 2.84&-1.56 &-0.199 \\      
Fe{\sc \,ii}\,$\lambda$2374 & 3.56&-2.01 &-0.254 \\      
Fe{\sc \,ii}\,$\lambda$2383 & 3.28&-1.86 &-0.234 \\      
Fe{\sc \,ii}\,$\lambda$2587 & 3.23&-2.16 &-0.250 \\      
Fe{\sc \,ii}\,$\lambda$2600 & 2.91&-1.97 &-0.227 \\      
Ni{\sc \,ii}\,$\lambda$1709 & 1.60&-0.47 &-0.082 \\      
Ni{\sc \,ii}\,$\lambda$1741 &-1.40& 0.42 & 0.073 \\
Ni{\sc \,ii}\,$\lambda$1751 &-0.60& 0.18 & 0.032 \\
Zn{\sc \,ii}\,$\lambda$2026 & 4.96&-2.04 &-0.301 \\      
Zn{\sc \,ii}\,$\lambda$2062 & 3.15&-1.34 &-0.195 \\\hline
\end{tabular}
\end{center}
\end{table}

Further examination of the $q_1$ coefficients in Table \ref{tab:atomdata}
reveals several other very useful combinations of lines. In particular,
note the large {\it negative} values for the Cr{\sc \,ii} lines. A
comparison between Cr{\sc \,ii} spectra and that of Zn{\sc \,ii} -- the
$\lambda$2026 transition having the largest magnitude value of $q_1$ --
provides the most sensitive combinations for probing non-zero $\da$. Also,
note the coefficients for Ni{\sc \,ii}. Here, within the same species (but
different multiplets), we have both positive and negative values for
$q_1$. This unique case is due to the very complicated multiplet structure
of Ni{\sc \,ii} as discussed in detail in Dzuba et al. (2001).

We may summarise the advantages of the MM method over the AD method as
follows:
\begin{itemize}
\item By including {\it all} relativistic corrections (i.e. including those
for the ground state) there is a sensitivity gain of around an order of
magnitude compared to the AD method.

\item In principle, all transitions appearing in a QSO absorption system
may be used.  This provides an obvious statistical gain and a more precise
constraint on $\da$ compared to using a single AD alone.

\item A further advantage of using many transitions is that the velocity
structure is determined with much greater reliability due to the larger
range of line strengths (see Section \ref{sec:spectra} for further
explanation).

\item A very important advantage is that comparison of transitions with
positive and negative $q_1$ coefficients minimizes systematic effects (see
Section \ref{sec:highz} in particular).
\end{itemize}

Finally, we note that the $q_1$ and $q_2$ coefficients for Al{\sc
\,ii}\,$\lambda 1670$ and Si{\sc \,ii}\,$\lambda 1526$ in Table
\ref{tab:atomdata} have not been presented elsewhere before. In the case of
Al{\sc \,ii}\,$\lambda 1670$, the coefficients were calculated using the
same methods described in Dzuba et al. (1999a,b). The coefficients for
Si{\sc \,ii}\,$\lambda 1808$ do appear in Dzuba et al. (1999b) but the
calculation has been re-performed to include other Si{\sc \,ii} lines,
including $\lambda 1526$. Thus, there is a small ($\approx 5\%$) difference
between the previous value and the one presented in Table
\ref{tab:atomdata}. This small difference is within the precision of the
Dzuba et al. (1999a,b) calculations and does not significantly affect our
determination of $\da$.

\section{Data requirements, acquisition and reduction}\label{sec:req}
The fitting of high-resolution (${\rm FWHM}\sim7{\rm \,kms}^{-1}$,
$R\sim45000$), moderate signal-to-noise ratio (${\rm SNR}\sim30$ per pixel)
QSO absorption lines routinely yields redshifts of single velocity
components measured to precisions of $\sigma(z)\sim 3\times10^{-6}$ or
$\sigma(v)\sim 1{\rm \,kms}^{-1}$ in velocity space (Outram, Chaffee \&
Carswell 1999). By comparison with the line shifts in Table
\ref{tab:shifts} and accepting that several well constrained velocity
components will exist in each absorption complex, the level of precision
achievable with the MM method is $\sigma(\da) \sim 10^{-5}$ per absorption
cloud. If systematic effects are to be avoided when using a large number
(i.e. $\sim 50$) of absorption systems then, from equation \ref{eq:omega},
we require $\omega_0$ to be known to a precision of $\sigma(\omega_0)\sim
0.003{\rm \,cm}^{-1}$ -- an order of magnitude more precise than available
wavenumber compilations (e.g. Morton 1991).

We therefore require high-resolution, moderate SNR QSO spectra and new
laboratory measurements of UV resonance transition wavelengths in order to
reach a precision of $\sigma(\da) \sim 10^{-5}$ for each absorption system.

\subsection{QSO Spectra}\label{sec:spectra}
\subsubsection{General properties}\label{sec:props}
All our QSO spectra were obtained at the Keck I 10\,m telescope with the
HIRES facility (Vogt et al. 1994). The spectra separate conveniently into
two samples with several defining features. One sample comprises spectra of
28 Mg{\sc \,ii}/Fe{\sc \,ii} absorbers in the spectra of 17 QSOs covering a
redshift range $0.5 < z_{\rm abs} < 1.8$. The second sample comprises 18
damped Lyman-$\alpha$ systems (DLAs) in the spectra of 11 QSOs covering
$1.8 < z_{\rm abs} < 3.5$. This later sample also contains three lower
redshift Mg{\sc \,ii}/Fe{\sc \,ii} systems (i.e. a total of 21 absorbers
comprises this sample). Different groups observed each sample. Thus, with
only a small degree of overlap, we label the samples according to the
typical redshift of the absorption systems. The major difference between
the \loz~and \hiz~samples is that the \loz~sample only contains transitions
of Mg{\sc \,ii} and Fe{\sc \,ii} whereas the \hiz~sample has a more diverse
range of transitions present (see Section \ref{sec:highz}).

\begin{figure*}\label{fig:mgfe}
\centerline{\psfig{file=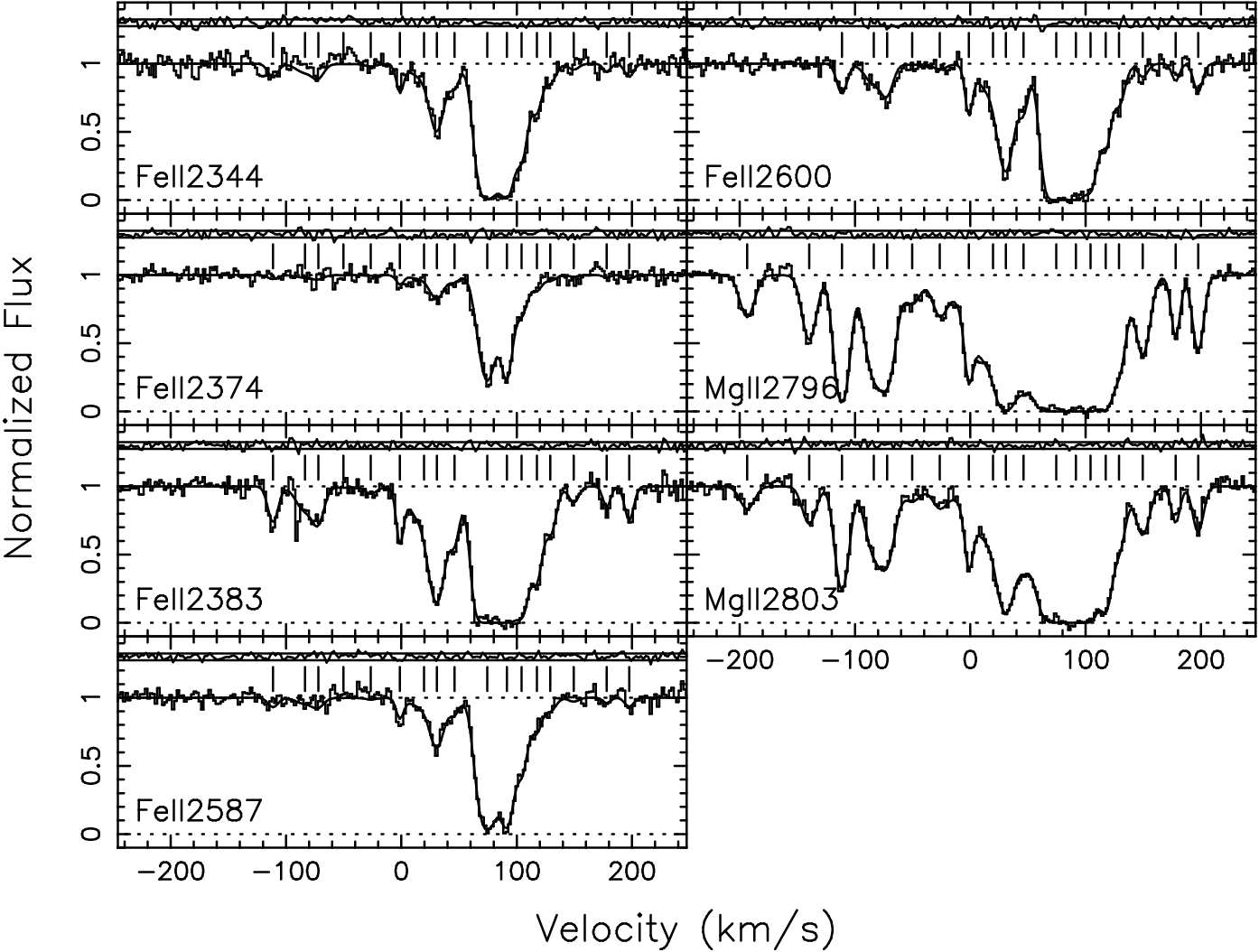,width=7in,angle=270}}
\caption{Mg/Fe absorption system towards Q1213$-$003 at $z=1.554$. The data
have been normalized by a fit to the continuum and plotted as a
histogram. Our Voigt profile fit (solid curve) and the residuals (i.e.
$[{\rm data}] - [{\rm fit}]$), normalized to the $1\sigma$ errors
(horizontal solid lines), are also shown. The tick--marks above the
continuum indicate individual velocity components.  Note the range of line
strengths in Fe{\sc \,ii}, facilitating determination of the velocity
structure. The large number of Fe{\sc \,ii} transitions and the large
number of velocity components allows for tight constraints to be placed on
$\da$. In this case, the Mg{\sc \,i}\,$\lambda2853$ is also present but is
sufficiently weak so as not to provide any constraint on $\da$. In cases
where the Mg{\sc \,ii} lines are more saturated, the Mg{\sc \,i} line is
typically stronger and can be used to constrain the velocity structure.}
\end{figure*}

\subsubsection{Low-$z$ ($0.5 < z < 1.8, \left<z\right> = 1.0$)}\label{sec:lowz}
This sample comprises the same data set analysed in W99, with some
differences as detailed in Section \ref{sec:analysis}. The observations
were taken in 1994 July, 1995 January and 1996 July. The SNR per pixel
ranged from 15--50 with most spectra having ${\rm SNR} \sim 30$. The FWHM
was $\sim$6.6${\rm ~kms}^{-1} (R=45 000)$. The QSOs were generally bright
($m_{\rm V} \la 17.5$) and so several short ($\sim$1000\,s) exposures were
taken of each object. The data were reduced with a combination of {\sc
iraf}\footnote{{\sc iraf} is distributed by the National Optical Astronomy
Observatories, which are operated by the Association of Universities for
Research in Astronomy, Inc., under cooperative agreement with the National
Science Foundation.} routines and personal code. Individual frames were
overscan subtracted, bias frame corrected and flatfielded. Cosmic rays were
removed by median filtering and the frames were averaged to form the final
image. 1$\sigma$ error arrays were generated assuming Poisson counting
statistics. We fitted continua to regions of each spectrum containing any
of the relevant transitions by fitting Legendre polynomials to
$\sim$500${\rm ~kms}^{-1}$ sections. Full details of the reduction process
are given in Churchill (1995, 1997) and Churchill et al. (1999). We have
also used the spectrum of Q0000$-$26 analysed by Lu et al. (1996).

Wavelength calibration of the frames was carried out within {\sc iraf}:
Thorium--Argon (ThAr) lamp exposures were taken before and after the QSO
exposures and co-added to provide a calibration spectrum. ThAr lines were
selected and centroided to provide a wavelength solution. The solution was
applied to the corresponding spectra without any re-sampling or
linearization so that the absorption line shapes were not distorted in the
process. We also note that no image rotator was installed during these
observations and so the spectrograph slit was not perpendicular to the
horizon in general. We make a detailed analysis of these points and their
effect on our results in M01b.

The transitions used in this sample are the five Fe lines, Fe{\sc \,ii}
$\lambda2344$--$\lambda2600$, and the Mg transitions, Mg{\sc \,i}
$\lambda2853$, Mg{\sc \,ii}\,$\lambda2796$ and $\lambda2803$. The Mg lines
have small $q_1$ coefficients and so act as anchors against which the
larger Fe{\sc \,ii} shifts can be measured. This set of lines is also very
useful because it offers a range of line strengths due to the differing
Fe{\sc \,ii} oscillator strengths (see Table \ref {tab:atomdata}) and,
together with the similar ionization potentials of Fe{\sc \,ii} and Mg{\sc
\,ii}, facilitates reliable determination of the velocity structure of the
absorption complex. In some cases the Mg{\sc \,ii} lines are saturated and
so the Mg{\sc \,ii} lines themselves provide only weak constraints on
$\da$. However, the weak Mg{\sc \,i}\,$\lambda2853$ then often serves to
provide a good constraint on the velocity structure -- a further advantage
of using Mg/Fe absorption systems to probe $\da$. To illustrate these
points, we provide an example absorption system, together with our Voigt
profile fits, in Fig. 1.

\begin{figure*}
\centerline{\psfig{file=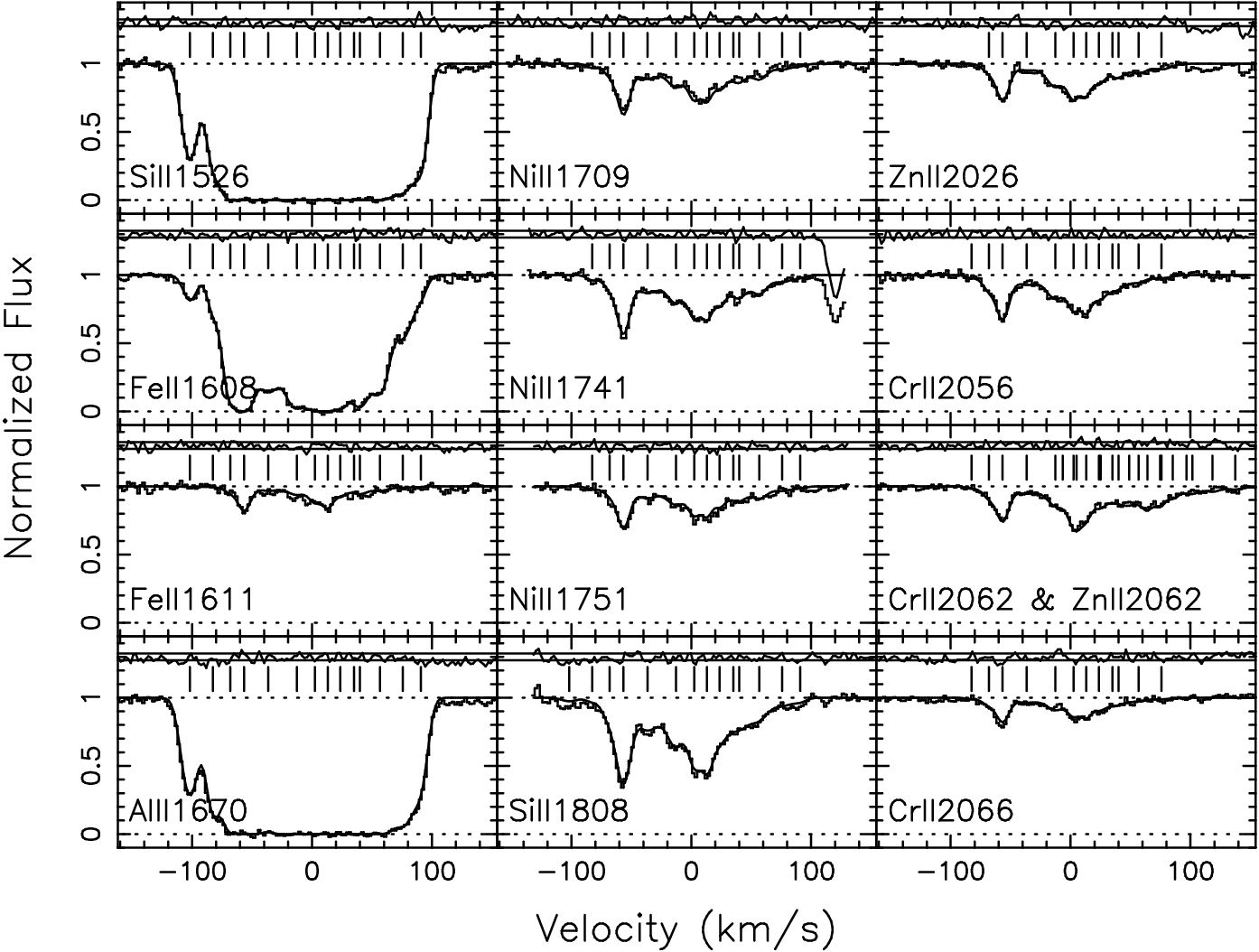,width=7in,angle=270}}
\caption{Heavy element absorption lines from the damped Lyman-$\alpha$
system towards Q2206$-$199 at $z=1.920$. We illustrate this system since it
displays both saturated and unsaturated transitions. However, the metal
lines in other DLAs generally have lower column densities. The data have
been normalized by a fit to the continuum and plotted as a histogram. Our
Voigt profile fit (solid curve) and the residuals (i.e.  $[{\rm data}] -
[{\rm fit}]$), normalized to the $1\sigma$ errors (horizontal solid lines),
are also shown. The tick--marks above the continuum indicate individual
velocity components. In this case, the Fe{\sc \,ii} and Si{\sc \,ii} lines
provide constraints on the velocity structure while the Ni{\sc \,ii},
Cr{\sc \,ii} and Zn{\sc \,ii} are strong enough to provide tight
constraints on $\da$. Note that the Cr{\sc \,ii} and Zn{\sc
\,ii}\,$\lambda2062$ profiles are blended together but that the constraints
on the velocity structure from all other transitions allow us to make a
statistically acceptable fit.}
\end{figure*}

\subsubsection{High-$z$ ($0.9< z < 3.5, \left<z\right> = 2.1$)}\label{sec:highz}
These observations were spread over many observing runs from 1994 to
1997. The SNR per pixel ranges from 15--40 with most spectra having SNR
$\sim 30$ and ${\rm FWHM}\la 7.5{\rm ~kms}^{-1}$ ($R = 34 000$). The QSOs
themselves are at higher redshift and so appear much fainter than those of
the \loz~sample: $m_{\rm V} \la 19.0$. Consequently, the total integration
times were much longer than for the \loz~sample and many more frames were
co-added during the reduction process.  An image rotator was installed in
1996 August and so only about half the observations were carried out with
the slit perpendicular to the horizon.
 
Most of the data were reduced using {\sc makee}, the HIRES data reduction
package written by T. Barlow. This package converts the two-dimensional
echelle images to fully reduced, one-dimensional, wavelength calibrated
spectra (calibration spectra were taken in the same way as for the
\loz~sample). Some of the spectra were reduced when {\sc makee} had no
wavelength calibration facility. In these cases, wavelength calibration was
carried out using {\sc iraf} routines. The remainder of the spectra were
fully reduced within {\sc iraf}.  1$\sigma$ error arrays were generated
assuming Poisson counting statistics. We fitted continua to regions of each
spectrum containing any of the relevant transitions by fitting Legendre
polynomials to $\sim 500{\rm ~kms}^{-1}$ sections. Full details of the
reduction procedures can be found in Prochaska \& Wolfe (1996, 1997,
1999). Outram, Chaffee \& Carswell (1999) have also kindly provided their
spectrum of Q1759$+$75 which was taken in 1997 July.

We also noted a small but important difference between the {\sc iraf} and
{\sc makee} wavelength calibration software. {\sc iraf} makes use of an
internal list of vacuum ThAr wavelengths and so spectra calibrated with
{\sc iraf} have ``correct'' wavelength scales. However, the {\sc makee}
package calibrates spectra with a list of air ThAr wavelengths and converts
the final QSO wavelength scale to vacuum using the Cauchy formula for the
refractive index (Weast 1979). However, this formula does not reproduce the
experimental dispersion: it shows deviations from experiment at the level
of $10^{-6}$ in the optical range. We therefore converted the wavelength
scales of the spectra reduced with {\sc makee} back to air wavelengths
using the Cauchy formula and then re-converted to vacuum wavelengths with
the preferred Edl\'{e}n (1966) formula. A detailed discussion of
air--vacuum conversion with caveats of the above statements is given in
M01b.

Theoretically, in damped systems such as those comprising this sample, the
combination of lines providing the best constraints on $\da$ are those of
Zn{\sc \,ii} and Cr{\sc \,ii}. In practice, however, these lines are
generally quite weak and so the parameter constraints are not
optimal. Also, the Zn{\sc \,ii} and Cr{\sc \,ii}\,$\lambda2062$ profiles
are often blended together (the transitions are only separated by $60{\rm
\,kms}^{-1}$), reducing the strength of the $\da$ constraint. The Ni{\sc
\,ii} lines are also generally weak but have the advantage of a variety of
widely differing $q_1$ coefficients. Also note that the only Fe{\sc \,ii}
lines available for these \hiz~absorbers are the Fe{\sc \,ii} $\lambda
1608$ and $\lambda 1611$ lines. These can be used in conjunction with the
two Si{\sc \,ii} lines to provide a reasonable constraint on $\da$ but
their main advantage is that they provide good constraints on the velocity
structure. We plot a typical absorption complex in Fig. 2 to illustrate the
above points and also to give the reader some idea of the relative
strengths of the different transitions used in analysing this sample.

In 6 spectra we have also fitted the two lines of Al{\sc \,iii}.
Variations in the incident radiation field could give rise to changes in
the relative column densities when comparing species with ionization
potentials above and below the Lyman edge. In our sample, the Al{\sc \,iii}
velocity structures corresponded well with those of the singly ionized
atoms (consistent with the conclusions of Wolfe \& Prochaska 2000) and the
value of $\da$ proved to be slightly better constrained by including the
Al{\sc \,iii} lines. In some cases, however, some velocity components in
the Al{\sc \,iii} profiles were much weaker or stronger relative to the
other components when compared with the profiles of singly ionized
species. In one case, our fitting algorithm removed many weak Al{\sc \,iii}
components from the fit (see Section \ref{sec:analysis}) and so we removed
the Al{\sc \,iii} data in this instance.

\subsection{Laboratory wavelength measurements}\label{sec:lab}
The laboratory rest wavelengths were measured by various groups using
Fourier transform spectrometry. We discuss the measurements for each ion
below, concentrating on those issues most relevant to our analysis. We
present the wavelengths that we have adopted in Table
\ref{tab:atomdata}. It is interesting to note that, with the exception of
the Fe{\sc \,ii} spectrum, the wavelengths presented in Table
\ref{tab:atomdata} are systematically longer than those compiled in Morton
(1991) -- a widely used list of resonance transition wavelengths. For a
particular transition, the wavelengths are generally consistent but, taken
altogether, a systematic trend is noted.

\begin{itemize}
\item Mg{\sc \,i} and Mg{\sc \,ii}: These spectra have recently been
measured by Pickering, Thorne \& Webb (1998). The spectra of Ni{\sc \,i}
and Ni{\sc \,ii} were excited along with those of Mg{\sc \,i} and Mg{\sc
\,ii} and several of these lines were used for calibration. The Ni{\sc \,i}
spectrum of Litz\'{e}n, Brault \& Thorne (1993) was used for reference
which itself was calibrated from the Fe{\sc \,i} and Fe{\sc \,ii} lines of
Nave et al. (1991). The Mg{\sc \,ii}\,$\lambda$2796 wavelength has been
measured previously by Drullinger, Wineland \& Bergquist (1980) and
Nagourney \& Dehmelt (1981) and the wavenumbers are in excellent
agreement. Also, a confirmation of the Pickering et al. (1998) wavenumbers
has been made and is described briefly in Pickering et al. (2000).

Since Mg is a relatively light atom, the isotopic structure of the line
profile must be taken into account. Pickering et al. (1998) used the
isotopic spacings measured by Hallstadius (1979) to obtain absolute
wavelengths for the isotopes. These values are listed in Table
\ref{tab:iso} and were used in our analysis.

\begin{table}
\begin{center}
\caption{The isotopic structures of the Mg and Si transitions. The
composite values are given in Table \ref{tab:atomdata}. The Mg isotopic
spacings were measured by Hallstadius (1979) but the spacings for Si are
estimates based on a scaling of the Mg{\sc \,ii}\,$\lambda$2796
structure. The hyperfine structure of $^{25}$Mg is also shown (Drullinger,
Wineland \& Bergquist 1980). The last column shows the relative abundance
of each isotope (Rosman \& Taylor 1998). Isotopic information was not used
for any heavier ions.}
\label{tab:iso}
\begin{tabular}{lcccc}\hline
Transition                 & $m/{\rm amu}$ & $\omega_0/{\rm cm}^{-1}$ &
$\lambda_0/{\rm \AA}$ & \%    \\\hline
Mg{\sc \,i}\,$\lambda$2853  & 24  & 35051.271  & 2852.9636   & 79.0\\
                            & 25  & 35051.295  & 2852.9616   & 10.0\\
                            & 26  & 35051.318  & 2852.9598   & 11.0\\
Mg{\sc \,ii}\,$\lambda$2803 & 24  & 35669.286  & 2803.5324   & 79.0\\
                            & 25  & 35669.300  & 2803.5313   &  2.7\\
                            & 25  & 35669.310  & 2803.5305   &  1.5\\
                            & 25  & 35669.360  & 2803.5266   &  2.8\\
                            & 25  & 35669.370  & 2803.5258   &  3.0\\
                            & 26  & 35669.388  & 2803.5244   & 11.0\\
Mg{\sc \,ii}\,$\lambda$2796 & 24  & 35760.835  & 2796.3553   & 79.0\\
                            & 25  & 35760.853  & 2796.3539   &  4.2\\
                            & 25  & 35760.913  & 2796.3492   &  5.8\\
                            & 26  & 35760.937  & 2796.3473   & 11.0\\
Si{\sc \,ii}\,$\lambda$1808 & 28  & 55309.330  & 1808.0132   & 92.2\\
                            & 29  & 55309.390  & 1808.0113   &  4.7\\
                            & 30  & 55309.446  & 1808.0094   &  3.1\\
Si{\sc \,ii}\,$\lambda$1526 & 28  & 65500.442  & 1526.7073   & 92.2\\
                            & 29  & 65500.517  & 1526.7055   &  4.7\\
                            & 30  & 65500.583  & 1526.7040   &  3.1\\\hline
\end{tabular}
\end{center}
\end{table}

\item Si{\sc \,ii}, Al{\sc \,ii} and Al{\sc \,iii}: These spectra have
recently been measured by Griesmann \& Kling (2000). The C{\sc \,iv} and
Si{\sc \,iv} spectra were also measured and all spectra were calibrated
using the Ar{\sc \,ii} spectrum of Whaling et al. (1995). At present, there
are no other similarly precise measurements of these spectra.

To our knowledge, the isotopic spacings for the Si{\sc \,ii} transitions
have not been measured. The mass shift and the specific mass shift should
dominate the volume shift in such a light ion. The $\lambda$1526 transition
is very similar to those of Mg{\sc \,ii} and so the specific mass shift
should be of the same order. Thus, to estimate the isotopic spacings in
Si{\sc \,ii}\,$\lambda$1526 we have scaled the spacings of Mg{\sc
\,ii}\,$\lambda$2796 by the mass shift: $\Delta\omega_i \propto
\omega_0/m^2$ for $\Delta\omega_i$ the wavenumber shift for a given isotope
$i$ where $m$ is the atomic mass. We have used these spacings to fit a
synthetic Gaussian emission line (centred at the composite wavelength) in
order to estimate the absolute wavenumbers for the isotopes. We present the
results in Table \ref{tab:iso}. We performed similar calculations for the
Si{\sc \,ii} $\lambda$1808 transition. However, the assumption that the
specific shift for $\lambda$1808 is similar to that for the Mg{\sc \,ii}
transitions is not so justified in this case since the transitions are of
quite different type (see Table \ref{tab:atomdata}). We discuss any effect
this approximation may have on $\da$ in M01b but find that it should be
negligibly small.

\item Cr{\sc \,ii} and Zn{\sc \,ii}: These spectra have been measured
independently using two different Fourier transform spectrometers: one at
Imperial College (IC), and another at Lund University (LU). The two
experiments are described in Pickering et al. (2000). The wavenumber
calibration was based on the Fe{\sc \,i} and {\sc ii} standards of Nave
et al. (1991) in both cases.

\item Fe{\sc \,ii}: This spectrum, from $1750$--$3850{\rm ~\AA}$, is
presented in Nave et al. (1991). The wavelength calibration was done by
comparison with the Norl\'{e}n (1973) Ar{\sc \,ii} spectrum. No other
(similarly precise) catalogues of Fe{\sc \,ii} lines exist in the
literature. We have obtained the Fe{\sc \,ii} $\lambda$1608 and
$\lambda$1611 wavenumbers from S. Johansson (private communication).

\item Ni{\sc \,ii}: This spectrum was recently measured by Pickering et al.
(2000). The original spectra were measured during the Mg{\sc \,i} and {\sc
\,ii} experiments by Pickering et al. (1998). Wavenumber calibration was
done using the Ni{\sc \,ii} standards of Litz\'{e}n, Brault \& Thorne
(1993) which cover the region down to about $2000{\rm \AA}$. The Ni{\sc
\,i} lines between $1750{\rm \AA}$ and $2000{\rm \AA}$ were calibrated with
the Fe{\sc \,i} and {\sc ii} lines so as to be used as calibration lines
for the Ni{\sc \,ii} spectrum. R. Kling (private communication) has
confirmed these wavenumbers, albeit with lower precision.
\end{itemize}

All wavelength calibrations above rely on the the Ar{\sc \,ii} spectrum
measured by either Norl\'{e}n (1973) or Whaling et al. (1995). However, the
Norl\'{e}n and Whaling calibrations systematically disagree. The Whaling
wavenumbers are larger and the difference between them is proportional to
the wavenumber: $\delta\omega = 7\times10^{-8} \omega$. Since this
difference is linear in frequency (velocity) space, as long as we
normalize all measured wavenumbers to the one calibration scale, any
systematic error will be completely absorbed into the redshift parameter
when we centroid corresponding velocity components in different transitions
in the absorption systems (see Section \ref{sec:analysis} for further
explanation). We choose to normalize all measured wavenumbers to the
Norl\'{e}n calibration scale and so the wavelengths of the Si{\sc \,ii},
Al{\sc \,ii} and Al{\sc \,iii} transitions in Table \ref{tab:atomdata} have
been scaled accordingly.

Also of note is that we only take into account the isotopic structures for
the lightest two elements, Mg and Si. In the optically thin limit, the
centroids of the isotopic structures described in Table \ref{tab:iso}
reduce to the composite values given in Table \ref{tab:atomdata}. When a
line becomes optically thick, differential saturation of the
isotopic components causes the centroid of the composite line to
shift. This effect is discussed in detail in M01b and it is evident that we
do not need to include isotopic information for the heavier ions.

\section{Analysis and Results}\label{sec:anal}
\subsection{Analysis}\label{sec:analysis}
Our analysis technique is based on a simultaneous $\chi^2$ minimization
analysis of multiple component Voigt profile fits to the absorption
features in several different transitions. Consider a QSO spectrum
containing a single velocity component of a specific transition. Three
parameters describe the Voigt profile fit to such a component: the column
density, the Doppler width or $b$-parameter and the redshift $z$ of the
absorbing gas cloud. In our case, however, each absorption system contains
many blended velocity components appearing in many different transitions.

The central assumption in the analysis is that the velocity structure seen
in one ion corresponds exactly to that seen in any other ion. That is, we
assume that there is negligible proper motion between corresponding
velocity components of all ionic species. With this assumption comes a
reduction in the number of free parameters to be varied in a particular fit
since the redshift of corresponding velocity components in different
transitions need only be specified once. We discuss this assumption and any
effect it may have on $\da$ in M01b.

Further restrictions can be placed on the number of free parameters when
fitting different transitions simultaneously since the $b$-parameters of
corresponding velocity components in different ionic species are physically
related. We may write
\begin{equation}\label{eq:b}
b_i^2 = \frac{2kT}{M_i} + b_{\rm turb}^2
\end{equation}
as the $b$-parameter of an ionic species $i$ which has a mass $M_i$. The
first term describes the thermal broadening within a gas cloud which has
a kinetic temperature, $T$, and the second term describes an additional
turbulent motion which affects all ions equally. If we assume that either
thermal or turbulent broadening dominates for a particular component then
again we reduce the number of free parameters. If we fit transitions of at
least two ions with different masses then we can determine $T$ and $b_{\rm
turb}$ for each velocity component.

For the present case, we wish to add another free parameter to the fit:
$\da$. From equation \ref{eq:omega} we can see that $\da$ will be
constrained by the difference $\omega_z - \omega_0$ for each transition if
we fit two or more transitions simultaneously\footnote{We require two or
more transitions since fitting only one would render $z$ and $\da$
degenerate for a single component and nearly degenerate for multiple
components.}.

We have used the program {\sc vpfit}\footnote{See
http://www.ast.cam.ac.uk/$^{\sim}$rfc/vpfit.html for details about
obtaining {\sc vpfit}.} (Webb 1987) to fit absorption profiles to the
spectra. Previously, {\sc vpfit} did not allow one to vary $T$ and $b_{\rm
turb}$ for each component independently, but dealt only with $b_i$.  There
was also no provision for including $\da$ as a free parameter. We have
therefore modified {\sc vpfit} to include $T$, $b_{\rm turb}$ and $\da$
explicitly as free parameters. This differs from the analysis in W99 where
$\da$ was varied externally to {\sc vpfit}. The main advantage is a
significant gain in computational speed, allowing us to fit more
complicated systems that would have taken prohibitively long using the
method in W99.  A minor point is that W99 presented 30 absorption systems
whereas we present 28 using the same QSO spectral data due to slightly
different subdivision of neighbouring absorption complexes.

Parameter errors can be calculated from the diagonal terms of the final
parameter covariance matrix (Fisher 1958). The reliability of these errors
has been confirmed using Monte Carlo simulations of a variety of different
combinations of transitions and velocity structures.

{\sc vpfit} imposes a cut-off point in parameter space such that very weak
velocity components are removed from the fit when they no longer
significantly affect the value of $\chi^2$. This presents a problem for our
method since a given component may be weak in one transition (and so can be
regarded as dispensable with respect to $\chi^2$) but may be strong in
another. An example of this can be seen in Fig. 2 where the two lowest
velocity components do not appear in the weaker transitions. Conceivably,
dropping even very weak components could affect our determination of $\da$.
Therefore, in such cases we observed the trend in the values of $(\da)_i$
at each iteration $i$ of the minimization routine to see if this trend was
significantly altered due to line dropping. If components were dropped
during a fit then we also re-ran the {\sc vpfit} algorithm, keeping the
dropped components by fixing their column density at the value just before
they were dropped from the original fit. No cases were found where the
values of $\da$ from the different runs differed significantly.

We impose several consistency checks before we accept a value of
$\da$. Firstly, the value of $\chi^2$ per degree of freedom must be
$\sim$1. Secondly, to check internal consistency, we make three different
fits to the data with three different conditions on the $b$-parameters: (i)
entirely thermal broadening, (ii) entirely turbulent broadening and (iii) a
fit where $T$ and $b_{\rm turb}$ are determined by goodness of fit. The
values of $\da$ derived from the three fits should be consistent with each
other since the choice above should not greatly affect $\da$ (i.e. the
redshifts of velocity components will not change systematically). If this
was not the case then the system was rejected. Only 2 systems were rejected
in this way. Since (iii) is the most physically realistic choice we expect
$\chi^2$ for that fit to be less than that for cases (i) and (ii). This was
indeed the case for all but a few systems. These exceptions were found to
have very low average temperatures (across the velocity structure) or very
small values of $b_{\rm turb}$. The difference between $\chi^2$ for the fit
for case (iii) and (ii) or (i) respectively was small in these cases. We
selected the final value of $\da$ from the regime which gave the lowest
$\chi^2$ but the results were insensitive to this choice.

\subsection{Results}\label{sec:res}
\begin{table}
\caption{The raw results from the $\chi^2$ minimization procedure. For each
QSO sight line we identify the QSO emission redshift, $z_{\rm em}$, the
absorption cloud redshift, $z_{\rm abs}$, the transitions utilized in our
analysis and the value of $\da$ for each absorption cloud.}
\label{tab:results}
\begin{center}
\vspace{-0.7cm}
\begin{tabular}{clllr}\hline
Object & $z_{\rm em}$ & $z_{\rm abs}$ & Transition$^a$ & $\da /10^{-5}$\\\hline
\multicolumn{5}{c}{\loz~sample}\\\hline
0000$-$26 &  4.11  &1.434$^b$&bcqr            &  $-0.732 \pm 2.283$\\
0002$+$05 &  1.90  & 0.851 &  bcnopqr         &  $-0.239 \pm 1.516$\\
0117$+$21 &  1.49  & 0.729 &  abcqr           &  $-0.403 \pm 1.298$\\
          &        & 1.048 &  bcnpr           &  $-1.171 \pm 1.800$\\
          &        & 1.325 &  bcpqr           &  $ 0.314 \pm 0.765$\\
          &        & 1.343 &  cnpq            &  $-0.982 \pm 0.868$\\
0420$-$01 &  0.915 & 0.633 &  abcr            &  $ 3.360 \pm 5.532$\\
0450$-$13 &  2.25  & 1.175 &  bnopr           &  $-2.590 \pm 0.831$\\
          &        & 1.230 &  bcnpqr          &  $-1.112 \pm 0.798$\\
          &        & 1.232 &  bcp             &  $ 0.611 \pm 2.475$\\
0454$+$03 &  1.34  & 0.860 &  acnopqr         &  $ 0.007 \pm 1.266$\\
          &        & 1.153 &  bcnpqr          &  $-0.340 \pm 2.047$\\
0823$-$22 &0.91$^c$& 0.911 &  bcnpqr          &  $-0.068 \pm 0.661$\\
1148$+$38 &  1.30  & 0.553 &  bcqr            &  $-1.882 \pm 1.699$\\
1206$+$45 &  1.16  & 0.928 &  bcnopqr         &  $-0.800 \pm 1.578$\\
1213$-$00 &  2.69  & 1.320 &  abcnopqr        &  $-1.821 \pm 0.893$\\
          &        & 1.554 &  bcnopqr         &  $-0.965 \pm 0.968$\\
1222$+$22 &  2.05  & 0.668 &  bcnpr           &  $-0.245 \pm 1.549$\\
1225$+$31 &  2.22  & 1.795 &  abcnopr         &  $-1.473 \pm 1.521$\\
1248$+$40 &  1.03  & 0.773 &  bcnopqr         &  $ 1.398 \pm 1.260$\\
          &        & 0.855 &  bcnpqr          &  $-0.329 \pm 1.301$\\
1254$+$04 &  1.02  & 0.519 &  abcqr           &  $-4.486 \pm 4.308$\\
          &        & 0.934 &  bcnpqr          &  $ 0.530 \pm 2.146$\\
1317$+$27 &  1.01  & 0.660 &  bcnpqr          &  $ 0.348 \pm 1.781$\\
1421$+$33 &  1.91  & 0.843 &  bcnopqr         &  $-0.111 \pm 0.793$\\
          &        & 0.903 &  bcnopqr         &  $-1.093 \pm 2.101$\\
          &        & 1.173 &  bcnpr           &  $-2.837 \pm 1.807$\\
1634$+$70 &  1.34  & 0.990 &  bcnpqr          &  $-1.586 \pm 3.752$\\\hline
\multicolumn{5}{c}{\hiz~sample}\\\hline
0019$-$15 &  4.53  & 3.439 &  ghl             &  $ 0.920 \pm 3.520$\\ 
0149$+$33 &  2.43  & 2.140 &  defghijklmstu   &  $-3.420 \pm 1.800$\\ 
0201$+$37 &  2.49  & 1.476 &  cnoqr           &  $-0.775 \pm 0.878$\\ 
          &        & 1.956 &  ehkl            &  $-0.444 \pm 3.243$\\ 
          &        & 2.325 &  deghl           &  $ 3.473 \pm 2.414$\\ 
          &        & 2.457 &  dgl             &  $ 0.902 \pm 3.754$\\ 
          &        & 2.462 &  ghkltu          &  $ 0.520 \pm 1.110$\\ 
0841$+$12 &2.20$^c$& 2.375 &  dghijtuvw       &  $ 0.285 \pm 3.670$\\
          &        & 2.476 &  dghijklmtu      &  $-1.351 \pm 3.052$\\ 
1215$+$33 &  2.61  & 1.999 &  defghijlmtuvw   &  $-0.972 \pm 3.411$\\ 
1759$+$75 &  3.05  & 2.625 &  eghklmstu       &  $-2.199 \pm 1.521$\\ 
          &        &2.625$^d$&dglmstu         &  $-0.354 \pm 1.613$\\ 
2206$-$20 &  2.56  & 0.948 &  bcnpqr          &  $-2.436 \pm 1.724$\\ 
          &        & 1.017 &  abcnopqr        &  $-0.372 \pm 0.689$\\ 
          &        & 1.920 &  dghijklmstuvw   &  $ 0.929 \pm 1.040$\\
2230$+$02 &  2.15  & 1.858 &  dghjlnprstu     &  $-1.615 \pm 1.699$\\ 
          &        & 1.864 &  ghijklmnostuvw  &  $-1.376 \pm 0.706$\\ 
2231$-$00 &  3.02  & 2.066 &  ghjlkmtuvw      &  $-1.590 \pm 1.265$\\ 
2348$-$14 &  2.94  & 2.279 &  ghl             &  $-0.244 \pm 5.204$\\   
2359$-$09 &  2.31  & 2.095 &  dghijklstuvw    &  $-0.946 \pm 0.601$\\ 
          &        & 2.154 &  dfgl            &  $-3.101 \pm 6.211$\\\hline 
\end{tabular}					       
\end{center}{\footnotesize
$^a$Transitions identified as in Table \ref{tab:atomdata}.\\
$^b$This absorber taken from Lu et al. (1996).\\
$^c$These QSOs are BL Lac objects and so their emission redshifts have
large uncertainties. Indeed, the literature values for $z_{\rm em}$ shown
here are less than the measured absorption redshifts.\\
$^d$This absorber contributed by Outram, Chaffee \& Carswell (1999).}
\end{table}

\begin{table*}
\caption{Statistics for the two subsamples and the sample as a whole. We
give the average redshift, $\bar{z}$, for each sample and the number of
data points, $N$, contributing to the weighted mean, $\left<\da\right>_{\rm
w}$, and unweighted mean, $\left<\da\right>$ (in units of $10^{-5}$). We
also give the significance, $\delta$, of the deviation from zero and the
reduced $\chi^2$ for the weighted mean. $\delta_{\chi^2_{\rm red}=1}$ is
the significance when we force $\chi^2_{\rm red}=1$ by scaling all errors
on $\da$ by $S$. The final two columns contain the results of the F test.}
\label{tab:stats}
\begin{center}
\begin{tabular}{lcccccccccc}\hline
Sample&$\bar{z}$&$N$&$\left<\da\right>_{\rm w}$&$\left<\da\right>$
&$\delta$&$\chi^2_{\rm red}$&$\delta_{\chi^2_{\rm
red}=1}$&$S$&$\sum_i(\chi^2_{\nalpha i}-\chi^2_{\alpha i})$&$\sum_i
\delta_i^2$\\\hline
Low $z$ &1.02&28&$-0.70\pm 0.23$&$-0.67\pm 0.27$&3.0$\sigma$&0.70&3.6$\sigma$&0.84&55&28\\ 
High $z$&2.12&21&$-0.76\pm 0.28$&$-0.67\pm 0.33$&2.8$\sigma$&0.68&3.3$\sigma$&0.83&37&21\\
Total   &1.49&49&$-0.72\pm 0.18$&$-0.67\pm 0.21$&4.1$\sigma$&0.68&4.9$\sigma$&0.83&92&49\\\hline
\end{tabular}
\end{center}			        
\end{table*}

\begin{figure*}
\centerline{\psfig{file=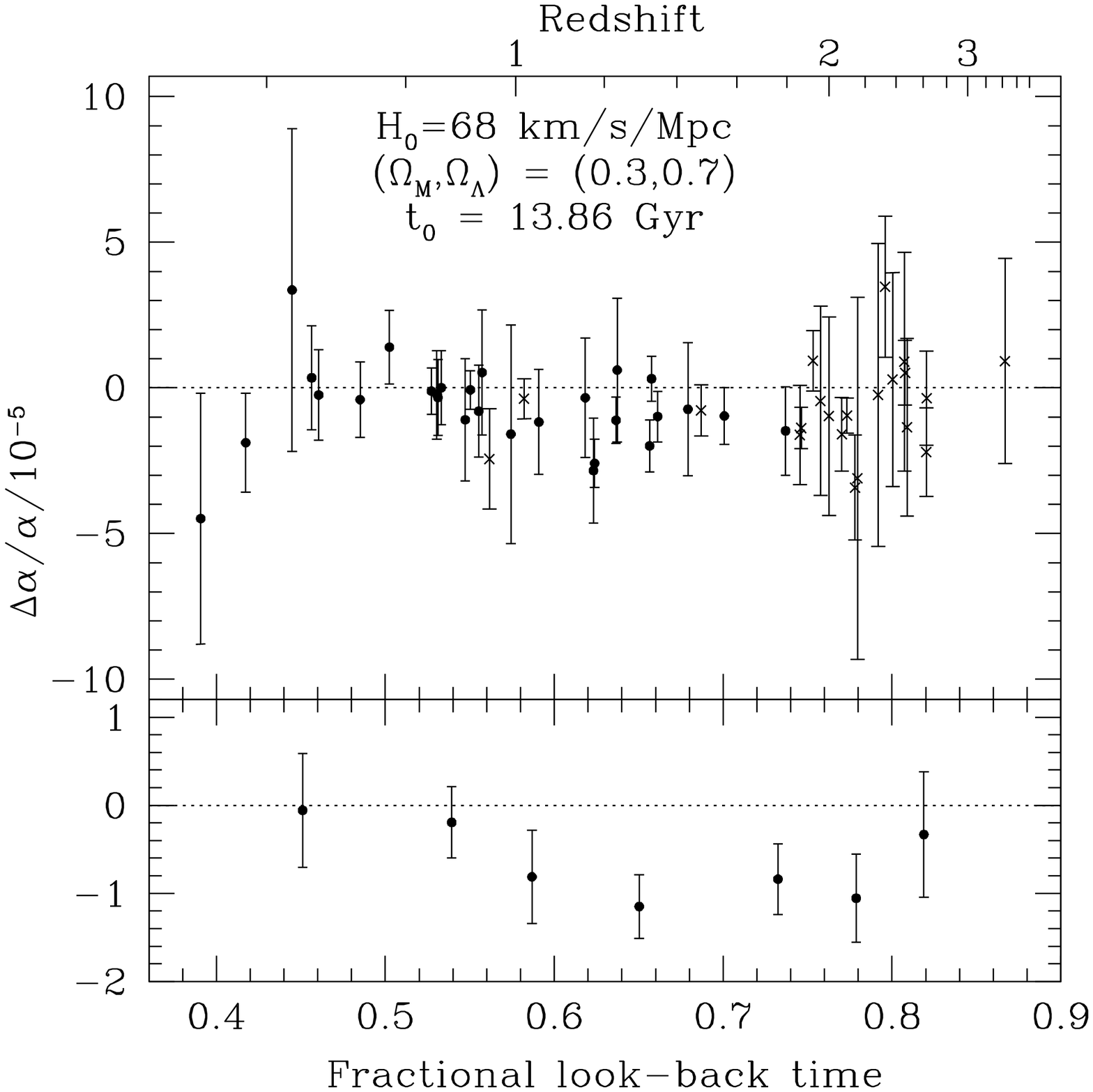,width=6.5in}}
\caption{$\da$ versus fractional look-back time for a currently popular
cosmology. The upper panel shows our raw results and $1\sigma$
error bars: the dots represent the \loz~sample and the crosses mark
the high redshift sample. Note that the \hiz~sample does contain some
lower redshift absorbers. The lower panel shows an arbitrary binning
of our results: 7 bins $\times$ 7 points per bin = 49 points. The
redshifts of the points are taken as the mean redshift of clouds
within that bin and the value of $\da$ is the weighted mean with its
associated $1\sigma$ error bar.}
\end{figure*}

We have analysed the \loz~sample in two stages. The first was motivated by
the fact that we have used different techniques to those of W99. We took
the final velocity structures for each absorption system found by W99, used
them as first guesses for the parameters and determined $\da$ with our new
methods. In all cases the results were nearly identical to those of
W99. The second stage was to fit the entire data set again, using no
information about previous velocity structures. These are the results we
present below.

Our results are presented in Table \ref{tab:results}. We give the object,
the QSO emission redshift, $z_{\rm em}$, the absorption cloud redshift,
$z_{\rm abs}$, $\da$ for that cloud and its associated 1$\sigma$ error. The
transitions for each cloud are indicated with the ID letters defined in
Table \ref{tab:atomdata}. Table \ref{tab:stats} shows the weighted mean,
$\left<\da\right>_{\rm w}$, the unweighted mean, $\left<\da\right>$ and the
significance level, $\delta$, of the weighted mean for the low and high
redshift samples and for the sample as a whole. We also include the value
of the reduced $\chi^2$, $\chi^2_{\rm red}$ (i.e. $\chi^2$ per degree of
freedom), for each sample where the model is taken to be a constant equal
to $\left<\da\right>_{\rm w}$.

Our results show that $\da$ deviates from zero at a significance level
of $4.1\sigma$ over the redshift range $0.5 < z < 3.5$. We also see a
general consistency between the low and \hiz~samples. The values of
$\chi^2_{\rm red}$ in Table \ref{tab:stats} are all less than
unity. This indicates that, if we have miscalculated the error bars
systematically, we have probably overestimated rather than
underestimated them. This is also reflected in a comparison of the
root-mean-square deviation from the mean, RMS, and the average size of
our error bars, $\left<\sigma(\da)\right>$: for the sample as a whole,
${\rm RMS} = 1.46 \times 10^{-5}$ and $\left<\sigma(\da)\right> = 2.01
\times 10^{-5}$. The RMS is also less than $\left<\sigma(\da)\right>$
for the low and \hiz~samples taken alone. Assuming we have
overestimated the errors on $\da$ then we can scale them by a constant
factor, $S$, so as to force $\chi^2_{\rm red}=1$. Column 8 in Table
\ref{tab:stats} shows the resulting increase in the significance of
$\left<\da\right>_{\rm w}$.  Note that the weighted means do not
differ significantly from the unweighted means for either sample
suggesting that the data error bars are reliable.

We have also conducted an F test on the $\da$ parameter. We used the final
velocity structures from the fitting procedure above to re-fit the
absorption systems after removing the $\da$ parameter. If $\da$ is
necessary to describe the data then we expect $\chi^2$ for each system to
reduce when we include $\da$. The reduction should be in accordance with
the statistical significance of $\da$ for each system.  We denote $\chi^2$
for the fit to system $i$ with and without the $\da$ parameter as
$\chi^2_{\alpha i}$ and $\chi^2_{\nalpha i}$ respectively.  If the
significance (in units of $\sigma$) of the value of $\da$ for that system
is $\delta_i$, we expect that
\begin{equation}\label{eq:ftest}
\sum_i^M(\chi^2_{\nalpha i}-\chi^2_{\alpha i}) = \sum_i^M \delta_i^2
\end{equation}
for $M$ absorbers. 

The $\delta_i^2$ can be derived from the last column of Table
\ref{tab:results}.  Direct calculation of both sides of equation
\ref{eq:ftest} yields the results in the final two columns of Table
\ref{tab:stats}.  $\sum_i(\chi^2_{\alpha i}-\chi^2_{\nalpha i}) \ga
\sum_i \delta_i^2$ for both the low and \hiz~samples, consistent with
$\da$ being a required parameter.

To illustrate the distribution of $\da$ over cosmological time, we plot our
results in Fig. 3. The upper panel of Fig. 3 shows the raw values of $\da$
as a function of fractional look-back time to the absorbing cloud using a
flat $\Lambda$ dominated cosmology ($H_0 = 68{\rm ~kms}^{-1}{\rm
Mpc}^{-1}$, $\Omega_{\rm M} = 0.3$, $\Omega_{\rm \Lambda} = 0.7$). The
redshift scale is also given for comparison. The lower panel shows an
arbitrary binning of the data such that all bins have equal number of
points (7 bins $\times$ 7 points per bin = 49 points). We plot the weighted
mean for each bin with the associated 1$\sigma$ error bars. At low
redshifts we see that $\da$ is consistent with zero. Also, the results are
consistent with a generally monotonic evolution of $\alpha$ with redshift.

\subsection{Systematic errors}\label{sec:syserr}
The statistical error in our result is now small and our attention must
turn to possible systematic errors.  In the companion paper, M01b, we
explore in detail a broad range of potential systematic errors, including
the following: laboratory wavelength errors, wavelength mis-calibration,
atmospheric dispersion effects, unidentified interloping transitions,
isotopic ratio and/or hyperfine structure effects, intrinsic instrumental
profile variations, spectrograph temperature variations, heliocentric
velocity corrections, kinematic effects and large scale magnetic fields.
All but two of these are shown to be negligible: atmospheric dispersion
effects and isotopic abundance variation.  In this Section we summarise
these two effects, but first discuss how a general wavelength scale
distortion affects the \loz~and \hiz~samples differently.

\subsubsection{Effect of a distortion of the wavelength scale}
To explain our results in terms of an effect other than real variation of
$\alpha$, such an effect must be able to mimic a non-zero average value of
$\da$ for both the high and \loz~samples alike. Generally, simple
systematic effects might cause low-order distortions in the wavelength
scale (relative to the ThAr calibration frames) or may lead to wavelength
dependent mis-centroiding of absorption features.

The \loz~sample (Mg{\sc \,ii}/Fe{\sc \,ii} systems) is most sensitive to
systematic effects. The Fe{\sc \,ii} lines have large and positive $q_1$
coefficients and all lie at lower wavelengths than the Mg{\sc \,ii}
anchors. Thus, a systematically negative $\da$ can result from a slight
`compression' of the QSO spectra relative to the ThAr calibration frames
for the \loz~sample. However, it is more difficult to understand the effect
of simple systematic errors on the \hiz~sample for several reasons:
\begin{enumerate}
\item The transitions involved have a large range of $q_1$ coefficients. In
effect, there are three types of transition: those with large and positive
$q_1$, those with large and negative $q_1$ and those with small or
intermediate values of $q_1$.

\item The different types of transition are mixed in wavelength space. In
the \loz~sample, all the lines with large and positive $q_1$ (i.e. Fe{\sc
\,ii} lines) lie to the blue of those with small $q_1$ (i.e. Mg{\sc \,ii}
lines). This is not the case in the \hiz~sample. For example, two lines
with small $q_1$, Si{\sc \,ii}\,$\lambda$1526 and Al{\sc \,ii}
$\lambda$1670, straddle two lines with large, positive $q_1$, Fe{\sc \,ii}
$\lambda$1608 and $\lambda$1611.

\item Different sets of transitions appear in each QSO spectrum. Some
spectra may not contain all types of transition.
\end{enumerate}

The value of $\left<\da\right>_{\rm w}$ will be more resistant to
systematic errors in the \hiz~sample. To illustrate this point we have
`compressed' the spectrum of a \loz~and a \hiz~absorption system by a
factor of $1\times 10^{-5}$ (i.e. the separation between any two
wavelengths decreases by 1 part in $10^{5}$) and found a new value for
$\da$. We used the same first-guess velocity structures used to derive the
results in Table \ref{tab:atomdata} and Fig. 3. For the \loz~Mg/Fe system
at $z=0.911$ we find a significant decrease in $\da$ due to the
compression: $\da = (-0.068 \pm 0.661) \times 10^{-5} \rightarrow (-1.106
\pm 0.692) \times 10^{-5}$. However, for the \hiz~DLA at $z=2.095$ we find
a small, {\it positive} change: $\da = (-0.946 \pm 0.601) \times 10^{-5}
\rightarrow (-0.821 \pm 0.589) \times 10^{-5}$.

\subsubsection{Atmospheric dispersion effects}\label{sec:atmodisp}
Atmospheric dispersion across the spectral direction of the spectrograph
slit can be avoided using an image rotator. However, as we noted in
Sections \ref{sec:lowz} and \ref{sec:highz}, many of our objects were
observed before HIRES was fitted with an image rotator. Dispersion across
the slit can only lead to a `stretching' of the QSO spectrum relative to
the ThAr frames. However, at least in the \loz~sample, a negative value of
$\da$ can only be mimicked by a `compression' of the spectrum. The effect
of the stretching on $\da$ is not quite as clear for the \hiz~sample
(points (i)--(iii) above).

We quantify the atmospheric dispersion effects in M01b and find that the
\hiz~sample is quite robust against this error. However, the \loz~results
are more sensitive to this effect and may have to be {\it reduced} from the
uncorrected value of $\left<\da\right>_{\rm w} = (-0.70 \pm 0.23) \times
10^{-5}$ to $\left<\da\right>_{\rm w} \approx (-1.52 \pm 0.23) \times
10^{-5}$. We stress that the magnitude of the correction here is an upper
limit since several effects, all difficult to properly quantify, will act
to reduce the effective compression. Note that since the above correction
is an upper limit, we do not include any additional errors due to
uncertainties in our model of atmospheric refraction.

\subsubsection{Isotopic abundance evolution}
We assumed terrestrial isotopic abundances in all species when fitting the
QSO spectra.  However, if the isotopic abundances in the QSO absorbers are
different, small apparent shifts in the absorption lines would be
introduced, potentially mimicking a non-zero $\da$.  $\da$ is most
sensitive to changes in the isotopic abundances of Mg{\sc \,ii} and Si{\sc
\,ii} since these ions have large isotopic separations (see Section
\ref{sec:lab}). Timmes, Woosley \& Weaver (1995), Timmes \& Clayton (1996)
and Gay \& Lambert (2000), suggest that the abundances of the isotopes
$^{25,26}$Mg and $^{29,30}$Si decrease with decreasing metallicity.  The
QSO absorption clouds in our sample have low metallicity compared to solar
(Prochaska \& Wolfe 1999, 2000; Churchill et al. 1999). Thus, to obtain an
upper limit on the effect this may have had on $\da$, we have removed the
$^{25,26}$Mg and $^{29,30}$Si isotopes from our analysis and re-calculated
$\da$ for all the absorption systems. Again, we find that the correction is
towards more negative values, the corrected value being $\da = (-0.96 \pm
0.17) \times 10^{-5}$ (as in Section \ref{sec:atmodisp}, we do not add any
additional errors due to uncertainties in our model since the correction we
calculate is an upper limit).

\subsubsection{Tests for unknown, simple systematic errors}
We have also carried out other tests to search for any unknown, simple
systematic effects. For example, it is unlikely that such an effect will be
able to mimic the very specific $q_1$ dependence of the transitions in the
\hiz~sample. Thus, if we remove the transitions with large, positive $q_1$
coefficients from our fit and find a new value for $\da$, we expect the two
values to differ if the line shifts are caused by some simple systematic
effect. Of course, if we remove transitions with large, positive $q_1$, we
can only obtain useful constraints on $\da$ if anchor lines {\it and}
transitions with large, negative $q_1$ are available. Therefore, we have
conducted such a test on a subset of the \hiz~sample which contains at
least one anchor line, at least one transition with large and positive
$q_1$ and at least one with large and negative $q_1$. We find consistent
weighted mean values of $\da$ both before and after line removal. We
conducted a similar test by removing lines with large, negative $q_1$ and
another test by removing the anchor lines. We find consistent values before
and after line removal in each case. Thus, the interpretation that the line
shifts we observe are caused by a varying $\alpha$ seems robust.

\section{Discussion}\label{sec:discussion}
We have conducted an experimental search for variability of $\alpha$ over
the redshift range $0.5 < z < 3.5$ using QSO absorption lines as our
probe. We used a new method for deriving constraints on $\da$ from a
comparison of QSO and laboratory spectra which, unlike the previous
alkali-doublet (spin-orbit splitting) method, exploits the total
relativistic effect in different atoms (see Section
\ref{sec:theory}). Comparing the wavelengths of transitions from different
multiplets and different atoms/ions leads to an order of magnitude gain in
precision over the alkali-doublet (AD) method. Since the new,
many-multiplet (MM) method allows us to use all transitions in principle,
we also gain precision through increased statistics.

We have analysed the spectra of 49 absorption clouds (towards 28 QSOs) and
compared the observed and laboratory wavelengths of those transitions
listed in Table \ref{tab:atomdata}.  Our \loz~results represent a
re-analysis of the W99 data. We compare the W99 results with our
re-analysis in Table \ref{tab:W99} and we see that they are consistent.
Our results for each absorption system are listed in Table
\ref{tab:results}. Over the entire sample we find statistical evidence for
time variation of $\alpha$ at the $4.1\sigma$ level: $\da = (-0.72 \pm
0.18)\times 10^{-5}$ (see Table \ref{tab:stats}).  This result is the `raw'
result.  Correcting for any possible systematic effects (Section
\ref{sec:syserr} and M01b) would {\it increase} the significance of this
deviation of $\da$ from zero.  We show the distribution of $\da$ with
cosmological time and redshift in Fig. 3.

\begin{table}
\caption{Comparison of results from W99 and the present work (\loz~sample
only) in different redshift regimes. Values of $\da$ are given in units
of $10^{-5}$.}
\label{tab:W99}
\begin{center}
\begin{tabular}{lccc}\hline
                  & \multicolumn{3}{c}{Redshift range}\\
                  & $z<1.0$       & $z>1.0$      & $0.5<z<1.8$\\\hline
Webb et al. (1999)& $-0.2\pm 0.4$ & $-1.9\pm 0.5$& $-1.1 \pm 0.4$\\
This paper        & $-0.2\pm 0.3$ & $-1.2\pm 0.3$& $-0.70 \pm 0.23$\\\hline
\end{tabular}
\end{center}
\end{table}

From the analyses we present in this paper and in M01b, we conclude that if
our results are due to some effect other than a real variation in $\alpha$,
this will only be revealed by further independent observations and analyses
using the MM method. Optical spectra are very useful for probing the range
of redshifts we have explored here. However, a statistical sample of radio
spectra of H{\sc \,i}\,$21{\rm \,cm}$ and molecular absorption systems has
the potential to increase our precision limit by another order of
magnitude. This effort is hampered by the fact that very few such systems
are known at present.  A systematic observational search for these systems
should clearly be made. Also, as the arsenal of $8$--$10{\rm ~m}$ class
optical telescopes with high-resolution spectrographs grows, more
observations should be devoted to obtaining carefully calibrated, high SNR
spectra of absorption systems at both high and \loz. These will provide a
crucial check on our results and will allow us to discover or rule out any
further subtle or unknown systematic effects in the data.

\section*{Acknowledgments}
We are very grateful to Ulf Griesmann, Sveneric Johansson, Rainer Kling,
Richard Learner, Ulf Litz\'{e}n, Juliet Pickering and Anne Thorne for
conducting laboratory wavelength measurements especially for the present
work and for communicating their results prior to publication. We would
also like to thank Michael Bessell, Tom Bida, Bob Carswell, Rolf Engleman
Jr., Alberto Fern\'{a}ndez-Soto, John Hearnshaw, Alexander Ivanchik, Jochen
Liske, Leon Lucy, Geoff Marcy, Gillian Nave and Steve Vogt for very helpful
communications.

We are grateful to the John Templeton Foundation for supporting this
work. MTM and JKW are grateful for hospitality at the IoA Cambridge, where
some of this work was carried out. CWC recieved partial support from NASA
NAG5-6399 and NSF AST9617185. JDB acknowledges support from PPARC, The
Royal Society, and a Gordon Godfrey visiting professorship at UNSW. AMW
received partial NSF support from NSF AST0071257.

\label{lastpage}
\end{document}